\pdfoutput=1
\RequirePackage{rotating}
\documentclass[acmsmall,nonacm]{acmart}

\usepackage{todonotes}
\usepackage{graphicx}

\usepackage{array} %
\usepackage{lscape} %
\usepackage{longtable} %
\usepackage[table]{xcolor}
\usepackage{wasysym} %

\AtBeginDocument{%
  }

\setcopyright{none}
\renewcommand\footnotetextcopyrightpermission[1]{}

\begin{document}

\title[Carbon-aware Resource Management for Latency-Sensitive Cloud Computing Environments]{Carbon-aware Resource Management for Latency-Sensitive Cloud Computing Environments: A Taxonomy and Future Directions}

\author{Tharindu B. Hewage}
\affiliation{%
  \institution{University of Melbourne}
  \city{Melbourne}
  \country{Australia}
}
\email{tsaryakarahe@student.unimelb.edu.au}

\author{Shashikant Ilager}
\affiliation{%
  \institution{University of Amsterdam}
  \city{Amsterdam}
  \country{Netherlands}
}
\email{s.s.ilager@uva.nl}

\author{Maria Rodriguez Read}
\affiliation{%
  \institution{University of Melbourne}
  \city{Melbourne}
  \country{Australia}
}
\email{maria.read@unimelb.edu.au}

\author{Patricia Arroba}
\affiliation{%
  \institution{Universidad Politécnica de Madrid}
  \city{Madrid}
  \country{Spain}
}
\affiliation{%
  \institution{University of Melbourne}
  \city{Melbourne}
  \country{Australia}
}
\email{p.arroba@upm.es}

\author{Rajkumar Buyya}
\affiliation{%
  \institution{University of Melbourne}
  \city{Melbourne}
  \country{Australia}
}
\email{rbuyya@unimelb.edu.au}

\renewcommand{\shortauthors}{Hewage, Ilager, Read, Arroba and Buyya}

\begin{abstract}

The proliferation of cloud-based latency-sensitive workloads, such as autonomous driving, real-time healthcare services, industrial automation, immersive interactive services, and intelligent AI-based applications requires infrastructures tuned to meet stringent latency constraints. Today, these workloads are reshaping the cloud from a general-purpose computing platform into a diverse ecosystem of workload-specific cloud environments. As demand for latency-sensitive workloads grows, cloud providers must scale their infrastructures accordingly, adversely increasing the carbon footprint and challenging net-zero goals. However, the strict performance requirements and rigid deployment patterns used to satisfy latency Service Level Objectives (SLOs) constrains conventional carbon optimization. Therefore, novel carbon-aware techniques are needed to exploit application-specific opportunities for reducing both operational and embodied carbon footprints while preserving latency guarantees. 
This survey examines carbon-aware resource management for latency-sensitive cloud computing environments. We introduce a holistic taxonomy that organizes existing techniques by carbon management objective, workload management mechanism, control architecture, system tier, latency requirements, and deployment stage. Then, we use this taxonomy to classify and critically analyze existing approaches, revealing the interactions and trade-offs between latency guarantees and operational and embodied carbon reduction. Finally, we identify limitations in the literature and present a structured research agenda for the design of sustainable latency-sensitive cloud systems.
\end{abstract}

\ccsdesc[500]{General and reference~Surveys and overviews}
\ccsdesc[500]{Computer systems organization~Distributed architectures}

\keywords{cloud computing, sustainability, de-carbonization, latency-sensitive applications}

\maketitle

\section{Introduction}

The primary difference between carbon-aware resource management for latency-sensitive workloads and carbon optimization in traditional hyper-scale cloud environments is the inability to compromise application workload performance for carbon efficiency. Cloud carbon optimization involves multiple dimensions during the life-cycle of cloud environments, from design and operation of data centers to termination of IT assets. Cloud systems are  are complex cyber-physical systems comprising  carbon-intensive stable energy sources, carbon-efficient variable renewable energy sources, IT assets that guarantee reliable and high performance through carbon-intensive frequent component upgrades due to aging and recycling of IT assets.
Consequently, carbon-aware resource management refers to the overall aspect of managing the resource requirements of an application workload while optimizing for carbon efficiency, emphasizing harmony between application performance and the intricacies of cloud environment elements. 
Traditionally, carbon-aware resource management techniques depend on spatio-temporal workload shifting algorithms between federated data centers with variable carbon intensity (CI). However, these techniques work with the assumptions that application workloads are tolerable to latency, thus making them only feasible to delay-tolerant batch workloads. However, due to the strictest service level objectives (SLOs) of latency-sensitive applications, there are fewer opportunities for carbon-aware resource management without compromising the performance requirements of application workloads. 

Latency-sensitive cloud workloads are increasingly prevalent across several critical domains~\cite{rasouli2025resource,batool2026ai}. Examples include autonomous driving and Advanced Driver Assistance Systems (ADAS), where perception and decision-making services must react within strict time bounds; healthcare applications such as remote patient monitoring, AI-assisted diagnostics, and robotic interventions; disaster management systems that support emergency coordination and situational awareness during the golden relief time following catastrophic events; industrial automation and digital-twin platforms supporting real-time control loops; 5G/6G network functions and edge-based telecommunications services; and interactive applications such as cloud gaming, augmented and virtual reality, and real-time AI assistants. In these application scenarios, even modest latency increases can degrade user experience, reduce operational efficiency, or, in safety-critical situations, compromise service reliability, disrupt critical infrastructure operations, and ultimately impact human safety. Consequently, carbon-aware resource management techniques cannot simply trade performance for carbon reductions, as is often possible in traditional cloud workloads, making the balance between sustainability and latency guarantees substantially more challenging.

This requires the exploration of complex application patterns intertwined with deployment architectures to uncover alternative means of carbon optimization while meeting SLOs. Consider, for example, an autonomous driving ecosystem in which cloud-edge services support cooperative perception, high-definition map updates, or vehicle-to-infrastructure coordination. Redirecting workload execution to a geographically distant site with lower CI may reduce emissions but can simultaneously introduce communication delays that affect the timeliness of safety-related decisions, potentially impacting the safe and dependable operation of transportation systems.

Similarly, personalized medicine platforms increasingly rely on cloud-hosted AI models to analyze patient-specific data and support time-sensitive clinical decision-making. While carbon-aware scheduling may favor execution in locations powered by cleaner energy sources, excessive delays can interfere with clinical workflows, reduce the usefulness of time-sensitive recommendations, and affect the timely delivery of critical healthcare services. In remote healthcare settings, including telemedicine and remote-assisted interventions, carbon optimization strategies must preserve stringent quality-of-service guarantees, as latency degradations may directly affect the effectiveness and safety of clinical procedures.

These examples illustrate the fundamental tension between carbon optimization objectives and strict latency requirements, where decisions aimed at reducing carbon emissions must be carefully balanced against performance guarantees that underpin service reliability, operational continuity, and, in some cases, public safety. We identify two significant aspects that must be addressed in a manner suitable for the particularities of latency-sensitive cloud environments.

\begin{itemize}
    \item \textbf{Operational Carbon Management:} As cloud environments operate, their IT assets consume electricity and produce heat as waste. This requires cooling systems to counterbalance the thermal energy. The energy consumption of IT assets and cooling systems draws power from various energy sources, which can have varying carbon intensities. The direct impact of the cloud environment on the carbon emissions of the energy source primarily defines its operational carbon footprint. As the carbon intensity of the energy source decreases, its supply variability increases more frequently, thus challenging application workload performance. Maintaining the required application latency performance while maximizing the integration of low-carbon intensity energy sources requires efficient resource management techniques.
    \item \textbf{Embodied Carbon Management:} Deploying, maintaining, and scaling physical elements of cloud environments incur an indirect carbon cost associated with carbon emissions made during the manufacturing, supplying, and recycling of IT assets, which are embodied in them, and generally categorized into scope-3 carbon emissions. As the demand for latency-sensitive applications increases, cloud environments must scale accordingly, while maintaining the performance of IT assets through frequent hardware refreshes, leading to the accumulation of embodied carbon. Providers must employ efficient resource management techniques to slow embodied carbon accumulation while maintaining the latency SLOs of application workloads by managing IT assets with heterogeneous reliability and performance characteristics.
\end{itemize}

Next, we identify the challenges associated with carbon-aware resource management techniques, which are significant for latency-sensitive workloads. We analyze these challenges from the perspectives of reducing the carbon footprint and meeting the latency SLOs of end users.

\begin{itemize}
    \item \textbf{Diverse latency requirements of application workloads:} Latency-sensitive application workloads exhibit varying degrees of latency tolerance in their SLOs. For instance, workloads may allow intermittent latency degradations, which end-user applications can tolerate. However, that may not be possible for time-critical applications, where SLOs expect strict bounds on the response time. Thus, these systems must apply carbon optimizations that adapt to the application latency requirements. Further, applications may exhibit different SLOs within their components. Navigating through these complex latency requirements in cloud environments is challenging, as impacting application SLOs can incur heavy penalties for the cloud operator.
    \item \textbf{Variable availability of carbon-efficient energy:} Often, energy sources with lower carbon intensity are renewable sources that are only available intermittently, such as solar and wind energy. Compared to stable sources, such as nuclear energy, which has low carbon intensity,  renewable energy plants with variable availability can be built and made operational at a lower cost and in less time. Additionally, energy capacity availability can vary spatially in geographically distributed cloud environments, which are becoming more prevalent due to their ability to execute  low-latency applications near end users. 
    Consequently, cloud environments face significant challenges in delivering consistent performance to meet application latency SLOs due to the intermittent availability of low-carbon energy capacities in both space and time. Failing to integrate intermittent energy sources could significantly increase the cloud environment's carbon footprint, while doing so must ensure that application performance remains intact.
    \item \textbf{Intermittent performance degradations in inter-cloud networks:} Cloud environments typically consist of hyper-scale data centers. A typical deployment pattern for latency-sensitive cloud applications is to use networked data centers that are geographically dispersed. These inter-cloud networks communicate through Wide Area Networks (WAN), which can undergo intermittent traffic congestion, degrading communication latency. Geo-spread availability of data centers enables cloud operators to redirect workload execution to match the dynamic availability of low-carbon energy capacity during the day. However, WAN traffic congestion challenges the opportunities to do so. Degraded network performance could increase traffic redirection delay, violating application latency SLOs.
    \item \textbf{Hard-constraints of resource allocation for latency performance:} Specific latency-sensitive applications require deterministic system performance to ensure critical service delivery. For instance, real-time services must meet hard bounds on their response times. In order to do so, they demand cloud servers to be tuned for high-performance and rigid allocation of computing resources, such as isolated CPU cores for virtual machines and turning off power optimization features of the CPU. As a result, the power management capabilities of the cloud environment are limited, especially when integrating low-carbon renewable energy sources with variable availability. Further, rigid resource allocations make redirecting workloads among sites with better carbon efficiency difficult due to limited placement opportunities satisfying the same placement constraints.
    \item \textbf{Resource-constrained cloud environments:} In order to satisfy the low-latency serving of application services, some latency-sensitive workloads must be deployed closer to users at the network's edge. These are primarily metropolitan areas where building new data centers can be expensive. A typical deployment pattern involves placing cloud servers in environments with resource constrains, such as colocation data centers, which allocate specific space and power budgets to servers. In return, the site's capability to utilize low-carbon energy sources or headroom with low-carbon servers is limited. Efficiently using the sites to reduce their carbon impact in resource-constrained environments becomes a complex task for the provider.
    \item \textbf{Degradation of IT assets:} Due to strict performance requirements, low-latency applications may use underlying hardware in an unsustainable manner, leading to premature degradation and carbon-costly frequency hardware replacements. For the cloud operator, alleviating hardware degradation is paramount for both cost and carbon. Since most hardware degradations are slow processes, understanding application patterns that lead to degradations over time and uncovering opportunities to optimize that while maintaining short-term application latency performance can be challenging.
\end{itemize}

Addressing the challenges discussed above requires a comprehensive understanding of how application requirements, infrastructure characteristics, resource management decisions, and carbon accounting dimensions interact in latency-sensitive cloud environments. Existing surveys provide valuable perspectives on carbon-efficient cloud computing, sustainable cloud infrastructures, and resource management for latency-sensitive applications. However, their scope and analytical focus vary, and the literature still lacks a unified characterization of how resource management decisions can jointly account for stringent latency requirements and the operational and embodied dimensions of carbon emissions. This survey addresses this need by structuring and synthesizing the relevant literature from the perspective of carbon-aware resource management for latency-sensitive cloud computing environments. The main contributions of this article are as follows: 

\begin{itemize} 
    \item \textbf{A holistic taxonomy of carbon-aware resource management.} 
    We propose a taxonomy that organizes the design space according to the main dimensions influencing carbon-aware resource management in latency-sensitive cloud environments. For operational carbon, the taxonomy characterizes existing approaches in terms of workload management mechanisms, control architectures, system tiers, and workload latency tolerance. For embodied carbon, it distinguishes resource management opportunities arising before and after infrastructure deployment. This structure provides a common basis for examining techniques that otherwise differ substantially in their objectives, assumptions, and deployment contexts. 
    
    \item \textbf{A taxonomy-driven classification and synthesis of existing techniques.} 
    We use the proposed taxonomy to classify and analyze the literature on carbon-aware resource management for latency-sensitive workloads. The resulting synthesis examines how existing approaches exploit workload flexibility, infrastructure heterogeneity, spatial and temporal variations in carbon intensity, renewable energy availability, and hardware lifecycle characteristics. It also identifies the application and infrastructure assumptions that determine the applicability of each class of techniques. 
    
    \item \textbf{An analysis of the interactions between carbon optimization and latency requirements.} 
    We examine how workload-specific latency requirements constrain carbon-aware resource management across different cloud and edge-cloud deployment tiers. In particular, we analyze the opportunities and limitations associated with workload placement, temporal and spatial shifting, resource allocation, throttling, preemption, hardware management, and other optimization mechanisms when latency SLOs must be preserved. This analysis also highlights the relationships and potential trade-offs between short-term operational carbon reductions and longer-term embodied carbon objectives. 
    
    \item \textbf{A research agenda for carbon-aware latency-sensitive computing.} 
    Based on the taxonomy and the limitations identified in the surveyed literature, we derive a set of open research directions for advancing carbon-aware resource management. These directions include spatio-temporal clean energy matching, hardware longevity and lifecycle-aware management, workload criticality and fault tolerance, data gravity, serverless runtime management, decentralized and multi-domain orchestration, long-term carbon budgeting, and robust AI-based predictive management. Together, they outline opportunities for reducing carbon emissions while preserving the performance and dependability requirements of latency-sensitive applications. 
    
\end{itemize} 

Overall, this survey provides a structured view of the current state of carbon-aware resource management for latency-sensitive cloud environments, the design decisions that characterize existing approaches, and the areas with substantial potential for further advances. The proposed taxonomy and literature synthesis can help researchers identify relationships and gaps across otherwise fragmented bodies of work, while supporting cloud operators and system designers in understanding the resource management options available for reconciling carbon reduction objectives with workload-specific latency requirements.

The rest of the article is organized as follows. An overview of the existing surveys and studies on carbon-aware resource management in latency-sensitive cloud environments is provided in Section \ref{sec:related-surveys}. Section \ref{sec:taxonomy} presents the proposed taxonomy of resource management. Section \ref{sec:classification-of-rm-techniques} summarizes existing works on carbon-aware resource management in latency-sensitive cloud environments based on the taxonomy. Finally, Section \ref{sec:research-gaps} provides an in-depth analysis of open research gaps and future directions, and Section \ref{sec:summary} concludes this article.

\section{Related Surveys}
\label{sec:related-surveys}

Literature surveys investigate the existing research in an area of interest to better understand its current state of progress. In that regard, many surveys study the carbon efficiency aspect of cloud environments. Preliminary surveys characterize carbon-efficient cloud computing as green cloud computing, emphasizing the integration of low-carbon intensity renewable energy sources \cite{kong14green-energy-dc, radu17green-cloud-computing,junaid16sus-cdc}. They study diverse aspects of resource management, such as workload scheduling, power management, and electronic waste management \cite{kong14green-energy-dc,radu17green-cloud-computing,junaid16sus-cdc}. 
Kong and Liu \cite{kong14green-energy-dc} categorized green-energy-aware datacenter power management across workload scheduling, capacity planning, and green energy harvesting. Similarly, Radu \cite{radu17green-cloud-computing} and Shuja et al. \cite{junaid16sus-cdc} reviewed resource management, waste heat utilization, and modular data center architectures to reduce carbon footprints. 
However, these classical surveys model computing infrastructures under generic workload abstractions (such as coarse batch or interactive jobs) and hyper-scale data center system architectures. Consequently, they concentrate almost exclusively on operational energy reduction, leaving real-time latency constraints and embodied carbon lifecycle trade-offs largely unaddressed.

Parallel to these developments, the state-of-the-art in latency-sensitive resource management has recently been synthesized by several comprehensive surveys focusing on predictability, time-criticality, and edge-cloud collaborative orchestration. For instance, Andreoli et al.~\cite{andreoli2025multi} provide a multi-domain survey analyzing how to support time-critical applications across the entire cloud software stack, spanning virtualization, orchestration, storage, and networking, to ensure temporal predictability. Similarly, Rasouli et al.~\cite{rasouli2025resource} systematically review edge resource allocation, offloading, and placement algorithms specifically designed for mission-critical applications where execution failures carry severe real-world consequences. In the context of distributed intelligence, Liu et al.~\cite{liu2026edge} comprehensively examine resource orchestration and model optimization techniques within heterogeneous edge-cloud collaborative computing architectures. Yet, despite their thorough treatment of strict latency bounds and multi-objective trade-offs, these modern surveys do not investigate carbon-aware resource management. They either treat sustainability as a generic energy-efficiency constraint to preserve device battery life~\cite{rasouli2025resource} or limit their green computing analysis to high-level renewable integration~\cite{liu2026edge}, overlooking the crucial distinction and trade-offs between operational and embodied carbon.

In contrast, more recently, a branch of surveys has emerged that studies carbon efficiency in cloud environments designed for emerging latency-sensitive workloads \cite{farahani2024towards,liu20strem-processing,souza24maintanance-iot,barrios23caching-edge,jeyaraj23em-cloud-iot,goudarzi22edge-iot,salaht20placement-edge,queiroz23cont-virt-rt}. In their studies, application latency-specific requirements are studied, such as low-latency streaming applications \cite{farahani2024towards,liu20strem-processing} and Internet of Things (IoT) applications \cite{souza24maintanance-iot,barrios23caching-edge,jeyaraj23em-cloud-iot,goudarzi22edge-iot,salaht20placement-edge}. 
Nevertheless, while these works rigorously tackle temporal predictability and service quality, they treat green computing either as an isolated energy-minimization metric \cite{liu20strem-processing, salaht20placement-edge} or as a secondary constraint for device battery longevity \cite{barrios23caching-edge, goudarzi22edge-iot, queiroz23cont-virt-rt}, omitting dynamic grid carbon intensity awareness and embodied carbon evaluation \cite{barrios23caching-edge, goudarzi22edge-iot, salaht20placement-edge, souza24maintanance-iot}.
This paradigm has evolved towards next-generation infrastructures, with recent studies advocating integrated, learning-centric approaches to cloud data center resource and workload management~\cite{buyya2024energy} while highlighting the multi-dimensional sustainability challenges faced by resource-constrained edge computing environments~\cite{arroba2024sustainable}.
Yet none of these studies consider the holistic view of carbon-aware resource management under latency-sensitive application execution. 
They consider carbon efficiency as one of many aspects of resource management and often overlook both operational and embodied carbon aspects. Given that power grids continue to decarbonize by integrating renewable energy sources, embodied carbon is more significant than ever. In this article, we address emerging latency-sensitive cloud applications through a holistic study of resource management techniques for both operational and embodied carbon management.

To systematically position our work within the state-of-the-art, Table~\ref{tab:related-surveys-comparison} provides a comprehensive taxonomic comparison of related surveys, evaluating them against the core dimensions of latency-criticality, operational and embodied carbon, and latency-sensitive holistic orchestration.
As highlighted in Table~\ref{tab:related-surveys-comparison}, a critical gap persists at the intersection of strict latency bounds, dynamic operational carbon awareness, and embodied carbon management. While early surveys focused on operational energy in centralized clouds and modern surveys address time-criticality in edge environments, this survey is the first to offer a holistic, carbon-aware resource management perspective tailored specifically for emerging latency-sensitive applications. By unifying these previously disjointed research vectors, this survey establishes a comprehensive taxonomy and strategic roadmap for sustainable, predictable, and next-generation latency-sensitive cloud environments.

\begin{table*}[t]
\centering
\caption{Taxonomic Comparison of Related Surveys with Respect to Carbon-Aware and Latency-Sensitive Orchestration}
\label{tab:related-surveys-comparison}
\footnotesize
\begin{tabular}{lcccc}
\toprule
\textbf{Survey Reference} & \textbf{\shortstack{Latency /\\Time-Criticality}} & \textbf{\shortstack{Operational\\Carbon}} & \textbf{\shortstack{Embodied\\Carbon}} & \textbf{\shortstack{Latency-sensitive\\Holistic Orchestration}} \\
\midrule
Kong \& Liu (2014) \cite{kong14green-energy-dc} & \LEFTcircle & \CIRCLE & & \\
Radu (2017) \cite{radu17green-cloud-computing} & & \CIRCLE & \LEFTcircle & \\
Shuja et al. (2016) \cite{junaid16sus-cdc} & & \CIRCLE & \LEFTcircle & \\
Farahani et al. (2026) \cite{farahani2024towards} & \CIRCLE & \CIRCLE & & \LEFTcircle \\
Liu \& Buyya (2020) \cite{liu20strem-processing} & \CIRCLE & \LEFTcircle & & \LEFTcircle \\
Souza et al. (2024) \cite{souza24maintanance-iot} & \LEFTcircle & \LEFTcircle & \LEFTcircle & \LEFTcircle \\
Barrios \& Kumar (2023) \cite{barrios23caching-edge} & \CIRCLE & \LEFTcircle & & \LEFTcircle \\
Jeyaraj et al. (2023) \cite{jeyaraj23em-cloud-iot} & \CIRCLE & \LEFTcircle & & \LEFTcircle \\
Goudarzi et al. (2022) \cite{goudarzi22edge-iot} & \CIRCLE & \LEFTcircle & & \LEFTcircle \\
Salaht et al. (2020) \cite{salaht20placement-edge} & \CIRCLE & \LEFTcircle & & \LEFTcircle \\
Queiroz et al. (2023) \cite{queiroz23cont-virt-rt} & \CIRCLE & \LEFTcircle & & \\
Andreoli et al. (2025) \cite{andreoli2025multi} & \CIRCLE & & & \LEFTcircle \\
Rasouli et al. (2025) \cite{rasouli2025resource} & \CIRCLE & \LEFTcircle & & \LEFTcircle \\
Liu et al. (2026) \cite{liu2026edge} & \CIRCLE & \LEFTcircle & & \LEFTcircle \\
Buyya et al. (2024) \cite{buyya2024energy} & \LEFTcircle & \CIRCLE & & \LEFTcircle \\
Arroba et al. (2024) \cite{arroba2024sustainable} & \CIRCLE & \CIRCLE & & \CIRCLE \\
\midrule
\textbf{This Survey (Ours)} & \CIRCLE & \CIRCLE & \CIRCLE & \CIRCLE \\
\bottomrule
\addlinespace
\multicolumn{5}{p{11.5cm}}{\small \textit{Notes:} \CIRCLE\ denotes broad discussion on the respective issue; \LEFTcircle\ denotes partial or conceptual discussion; blank spaces indicate no coverage.}
\end{tabular}
\end{table*}

\section{The Taxonomy}
\label{sec:taxonomy}

\begin{sidewaysfigure}
    \centering
    \includegraphics[width=0.9\linewidth]{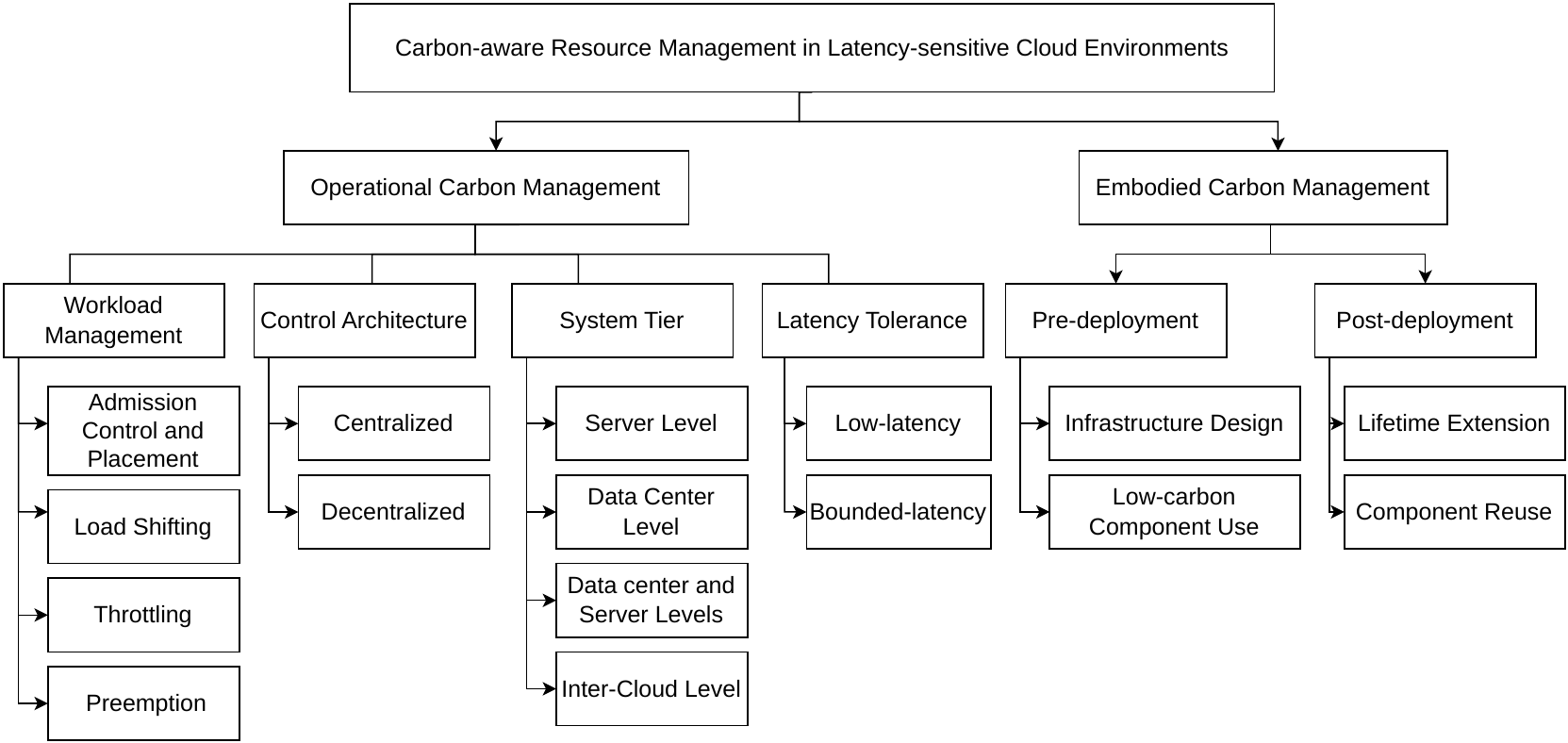}
    \caption{The taxonomy of resource management in serverless computing environments.}
    \label{fig:taxonomy}
\end{sidewaysfigure}

At the topmost level, our taxonomy categorizes carbon-aware resource management into the two primary aspects that we identified: operational and embodied carbon management. Operational carbon management comprises workload management, control architecture, system tier, and latency tolerance. Embodied carbon management comprises pre-deployment and post-deployment stages. Figure \ref{fig:taxonomy} illustrates the proposed taxonomy. In the following sections, we discuss each category in detail, referring to techniques explored in the literature so far.

\subsection{Operational Carbon Management}

Most works on carbon efficiency focus on reducing carbon emissions at the operational level of the cloud environment. 
While response latency serves as the primary SLO constraint in latency-sensitive cloud environments, system throughput (measured in transactions per second or processed tasks per $gCO_2eq$) plays a crucial complementary role in determining overall carbon efficiency. 
Maximizing computational throughput during low-carbon grid windows allows cloud providers to achieve higher work output per unit of carbon emitted ($gCO_2eq/\text{FLOP}$). 
Quantitative metrics such as Carbon Usage Effectiveness (CUE), real-time carbon intensity ($gCO_2eq/\text{kWh}$), and the Latency-Carbon Product (LCP) provide standardized benchmarks to evaluate the practical efficacy of operational trade-offs across heterogeneous cloud-edge tiers.

In this context, the foundational approach to improving such metrics is utilizing low-carbon-intensive energy sources to power the IT assets. In doing so, resource management techniques are challenged by the variable availability of the renewable energy sources present today, such as solar and wind. Operators opt for variable-availability renewable energy plants due to the lower cost and deployment times, in contrast to low-carbon-intensive energy plants that provide relatively stable electricity generation. For instance, nuclear and hydro plants capable of providing stable, low-carbon-intensive energy can take many years to finish construction, whereas a solar power plant can be deployed in a few years. Adversely, cloud operators integrating renewable energy sources must match their variable availability with the data center power load. This introduces significant challenges in ensuring the timing constraints of latency-sensitive workloads over data center load adjustments. Therefore, resource management techniques must address the integrated management of application performance and data center power load. This section identifies and briefly reviews their methods under the key resource management aspects of operational carbon management.

\subsubsection{Workload Management}

Workload execution yields dynamic power consumption patterns in cloud servers. In clouds, workload execution is carried out in isolated environments from the cloud operator due to privacy concerns, leaving the operator with high-level abstractions of running workloads. In return, the operator must utilize those abstractions to match the server load to the dynamic availability of renewable energy sources. In the following sections, we discuss prominent methods present in the literature, emphasizing how existing works leverage them for latency-sensitive workloads.

\par\noindent\textbf{Admission Control and Placement:}
Distributed system architectures of cloud environments typically consist of multiple data centers or compute clusters that provide heterogeneous performances for latency and carbon efficiency. Therefore, cloud providers have the opportunity to optimize carbon efficiency via the admission control or placement decisions made upon the arrival of workloads. Workloads are scheduled to execution sites, improving renewable energy utilization with minimal impact on latency performance.

Yuan et al. \cite{yuan2019delaytolerantgeodatacenters} study the problem of scheduling tasks among distributed green data centers to meet their response time constraints. Data centers in their system model integrate on-site renewable energy and periodically relay energy metrics to a centralized task scheduler, such as the electricity price of the power grid and the conversion rate of wind and solar radiation to electricity. Users' tasks arrive at the centralized scheduler and are queued according to the application to which each task belongs. They propose a task scheduling algorithm to split the queued tasks among distributed green data centers to minimize the combined energy cost of both the power grid and renewable energy while strictly meeting the task's delay-bounded constraints. In contrast to centralized task arrival, tasks can also arrive in each compute region. Chien et al. \cite{chien23genai-rq-direct} design request direction algorithms to minimize the carbon cost of latency-sensitive generative AI inference requests. Their system model consists of multiple compute regions with different carbon intensities. When a compute region receives a user request, it will be directed to the region with minimum carbon intensity, where the placement decision depends upon the calculated latency of both computation and round trip network time for each compute region. Admission control and placement also fit well with workloads with bounded response times, such as real-time services, because carbon efficiency can be estimated prior to service deployment. Kaur et al. \cite{kaur20keids-rt} investigate scheduling containers of real-time services in the edge-cloud continuum. They schedule containers among Kubernetes clusters having on-site renewable energy integration. Their scheduling algorithm encompasses carbon footprint minimization and constraints to maintain the real-time performance of services, such as reducing network interference.

Recent advances extend carbon-aware placement to multi-tier, federated, and IoT environments. 
Al-Shammaa and Al-Hadrawi \cite{ALSHAMMAA2026101398} present a carbon-aware federated orchestrator for multi-domain edge-cloud systems that employs dynamic adaptive weighting to balance carbon emissions, energy costs, Service-Level Agreement (SLA) compliance, and resource utilization while strictly respecting data sovereignty and domain trust boundaries. 
In hierarchical IoT architectures, Gao et al. \cite{11407972} propose the CECO algorithm, which uses Lyapunov optimization to solve dynamic offloading and computing resource allocation across cloud-edge-device tiers, taking advantage of green energy harvesting at edge servers. 
Similarly, Jiang et al. \cite{11666925} introduce CARBONMOBILE for mobile IoT edge offloading, exposing fine-grained per-edge action spaces to exploit intra-city grid carbon intensity variations while enforcing long-term carbon budgets via a Lyapunov virtual queue coupled with deep reinforcement learning.

\noindent\textbf{Load Shifting:}
Load shifting involves either suspending or resuming workloads (i.e., shifting in time) or geographically migrating workloads (i.e., shifting in space). Both aim to match the abundance of renewable energy, which varies based on time or location. Load shifting with latency-sensitive workloads primarily leverages shifting in space to avoid the latency violation risks of suspending workloads. Shifting in space involves chasing renewable energy availability across geographical locations, in which network latency performance and scheduling overhead must be efficiently managed. 

Sajid et al. \cite{sajid2021blockchain} propose connecting geographically distributed data centers with high-performance networks and addressing scheduling overhead with a blockchain-based decentralized approach. Sun et al. \cite{sun24modular-datacenters} leverage space shifting but use it as a last resort to incur minimal impact on workloads. In return, they achieve an average space shifting of 0.015 times per VM per hour, minimizing the associated latency overhead. For workloads that allow for tolerable latency slack, Sukprasert et al. \cite{sukprasert24limits-of-load-shift} conduct space shifting within a subset of locations that incur a latency impact within the tolerable slack. 
Limiting the use of a subset of locations can reduce overall carbon efficiency. To mitigate this issue, they propose complementing existing workloads with shiftable batch workloads. Murillo et al. \cite{murillo24cdnshifter} exploit tolerable latency slack in content delivery networks (CDNs) and show that in CDNs, increasing the latency slack to 60 ms can lead to over 60\% reduction in carbon emissions. To do so, they complement space shifting with a second form of shifting, called capacity shifting. Capacity shifting moves CDN capacity to greener regions at a coarser time scale of days or weeks. Another opportunity lies in the workload portability present within applications. Gsteiger et al. \cite{sun24modular-datacenters} exploit space-shifting opportunities within serverless applications. They leverage functions outside the application workflow's critical path and use them to perform space shifting. As a result, end-to-end application latency is left intact. Space-shifting opportunities are also present in specific application load-balancing scenarios. Souza et al. \cite{souza24casper} exploit space-shifting opportunities in distributed web services. They conduct dynamic server provisioning and load balancing for dynamic carbon intensities and network latency constraints across geo-distributed regions.

For complex big data and mobile workloads, load shifting must balance temporal renewable availability with spatial overheads. 
Bhat et al. \cite{11434192} present Carbon-Kube, a Kubernetes-native scheduler for big data Directed Acyclic Graph (DAG) pipelines (e.g., Apache Spark) that leverages forecast-aware temporal planning to defer batch jobs to low-carbon windows, explicitly incorporating data gravity penalties to prevent costly cross-region migrations. 
In ultra-low latency 6G Computing Power Networks (CPNs), Batool et al. \cite{batool2026ai} propose ProactiveCPN, an AI-native predictive framework utilizing multi-scale machine learning (LSTM, GNN, and Transformers) for proactive spatial positioning and mobility-aware resource allocation, shifting resources before demand peaks materialize.

\noindent\textbf{Throttling:}
Throttling is an in-place technique to dynamically adjust server performance without shifting the running load. Using throttling, the cloud operator can degrade the servers' performance to reduce their power draw, matching the variable availability of renewable energy. In return, servers take a prolonged time to execute workload instructions, impacting workload service quality, such as increasing the response latency in latency-sensitive applications. Therefore, throttling must be carefully orchestrated so that its impact on the application does not violate SLOs.

Li et al. \cite{li2013chameleon} use a data center power delivery architecture that provides access to both renewable energy sources and energy storage and propose a CPU hardware controller called Chameleon. Chameleon switches between two modes of power management to match renewable energy dynamics: an energy mode that throttles server performance using dynamic voltage and frequency scaling and a performance mode that uses stored energy for power deficiencies. When applied, DVFS degrades CPU performance, affecting running workloads. The authors employ a reinforcement learning-based control mechanism to efficiently manage mode switching to mitigate this issue. Jahanshahi et al. \cite{jahanshahi2022powermorph} propose a data center power shaping framework that leverages throttling through DVFS to match the dynamics of renewable energy. They co-locate latency-critical workloads with latency-tolerant complementary workloads. Then, CPU cores are grouped into three categories: working, offset, and free. Latency-critical workloads are then pinned to working cores with minimal throttling possibilities, and the remaining cores are allocated for complementary workloads. With that, they leverage data center-level power shaping signals to manage core groups efficiently, balancing throttling and workload performance.

Sharma et al. \cite{sharma11blink} propose an application-independent power management mechanism called Blink. Blink activates and deactivates servers to yield a duty cycle and, as a result, it can control the average power consumption of servers. Applications that can leverage the Blink architecture modify their design to adopt it. The duty cycle in Blink throttles servers, which is specified through a blinking policy. The blinking policy balances power management and application performance depending on the application. Experiments with latency-sensitive applications such as Memcached show that blinking can be adopted with only a modest throttling overhead. Govindan et al. \cite{govindan11stored-energy-dc} experiment with battery energy storage to utilize renewable energy and use application throttling to sustain battery longevity via less frequent discharge cycles. By combining both, they aim to balance application throttling and premature battery failures. Agarwal et al. \cite{ambati2020harvest-vms} propose application throttling through host virtualization to match renewable energy variations. They dynamically change the number of physical cores mapped to virtual machines while keeping the number of virtual cores intact. In return, application interruptions are avoided, yet CPU performance can be throttled. Their approach applies to applications that can tolerate a specific degree of degradation in latency performance. Souza et al. \cite{souza23ecovisor} propose a virtualized data center-level energy system to allow applications to control the energy sources according to their workload patterns. In return, applications can better adapt to renewable energy dynamics by exploiting application-specific usage patterns. Here, applications are provided with APIs to throttle their application containers through power capping to control their power draw, which is then exercised to better match their latency SLOs.

In microservice and serverless paradigms, throttling and resource scaling are dynamically adjusted to align with carbon signals without violating performance guarantees. 
Almaipeta et al. \cite{11542095} propose CARC, a Kubernetes-integrated controller that dynamically scales microservice replicas and defers non-critical background tasks based on real-time carbon telemetry and SLO risk models. 
For serverless computing, Sun et al. \cite{11619171} introduce LACE-RL, a deep reinforcement learning framework that dynamically tunes function pod keep-alive durations, balancing cold-start latency penalties against idle pod carbon emissions under time-varying grid carbon intensity.

\noindent\textbf{Preemption:}
Isolated execution environments in clouds typically do not reveal underlying infrastructure dynamics to users. For example, a virtual machine's SLOs are to provide a dedicated and stable physical machine, regardless of underlying resource allocation dynamics. Adversely, cloud operators lose the opportunity to offload infrastructure dynamics to end users where applicable, such as absorbing the variable availability of renewable energy. However, recent cloud offerings, such as preemptible VMs \cite{google22spot-vms,microsoft20spot-vms}, attempt to narrow this by providing virtual machines that reflect infrastructure-level dynamics to users. Preemptible VMs unlock the opportunity to relay variable availability of renewable energy to the application layer in specific latency-sensitive cloud environments, given that the application layer ensures fault tolerance. Preemption allows for reducing server load upon the loss of renewable energy capacity through fault tolerance.

Sun et al. \cite{sun24modular-datacenters} leverage preemptive VMs to delay drawing power from the carbon-intensive grids in geographically distributed networks of modular data centers integrating renewable energy. Their application layer allows users to define fault tolerance by deploying both preemptive and regular VMs. In case of a loss of energy capacity, they first attempt to reduce the data center power draw via VM migrations. If that is insufficient, they opt to shut down preemptive VMs to further reduce the data center power draw rather than sourcing energy from the carbon-intensive power grid. In doing so, an up-time threshold is maintained for preemptive VMs, and VM shutdowns are conducted to adhere to that.

\subsubsection{Control Architecture}

Integrated management of workloads and variable availability of renewable energy sources requires coordination between different resource management elements such as workloads, data centers and servers, and power delivery. In that regard, there are two prominent control architectures: centralized and decentralized. In this section, we discuss how related works apply their chosen control architecture while adhering to the latency constraints of their cloud environment design.

\noindent\textbf{Centralized:}
A centralized architecture enables a single point of control, providing ease of management to the cloud operator. For instance, it allows operators to monitor the cloud environment and execute scheduling logic centrally, yielding better maintenance of scheduling logic, such as extending scheduling logic for future requirements. Furthermore, system elements such as servers and data centers in centralized control are synchronized by design, eliminating the need to perform complex state synchronizations.

Yuan et al. \cite{yuan2019delaytolerantgeodatacenters} leverage centralized control to schedule tasks for applications replicated across multiple data centers that integrate on-site renewable energy. They employ per-application queues and execute scheduling logic to optimize green energy while meeting delay-bound constraints. Jahanshahi et al. \cite{jahanshahi2022powermorph} design PowerMorph: a power reshaping framework to support the frequency regulation requirements of power grids that can integrate renewable energy sources. PowerMorph reshapes the power consumption of data centers that execute latency-critical applications. A critical challenge in their design is that frequency regulation bids happen every hour, yet the data center should follow the regulation signal every two seconds, where the latter does not provide enough opportunity to perform cluster-level optimization. To overcome that, they offload it to the server level and utilize centralized control at the data center level to conduct regulation provisions. 

Sharma et al. \cite{sharma11blink} propose an application-independent centralized control plane that manages cluster power. Through the APIs, latency-sensitive applications interact with the control plane to regulate their power consumption through opportunities present in specific application patterns. In return, the cluster can integrate renewable energy sources while offloading load adjustment to applications. Govindan et al. \cite{govindan11stored-energy-dc} introduce energy buffers using batteries to integrate renewable energy in data centers. They use a centralized peak power budget enforcer component to leverage a hybrid approach combining batteries and server throttling. The hybrid approach performs better in dynamically adapting workload service level agreements, such as latency constraints. Sun et al. \cite{sun24modular-datacenters} use a centralized VM scheduler to obtain the global perspective of a modular data center network that integrates on-site renewable energy. It then uses that to find the best fit for VMs to achieve optimized decisions across a mix of modular data centers, priority VMs, and non-priority VMs. In return, the least number of interruptions are made with priority VMs, such as VMs executing latency-sensitive workloads. Murillo et al. \cite{murillo24cdnshifter} conduct load shifting in content delivery networks deployed on geographically distributed edge data centers that integrate renewable energy. They implement a centralized management component that provides an information service for weather and energy costs, capacity, and demand. As a result, the central component makes informed decisions about spatial load balancing and capacity shifting, considering trade-offs between latency, carbon, and cost. Gsteiger et al. \cite{gsteiger24caribou} leverage centralized carbon forecasting, pricing, and transmission latency to offload serverless workflows across geo-distributed regions with different carbon intensities. Through centralized management, they identify and deploy serverless functions that have the potential to offload while keeping latency-sensitive functions intact. Similarly, Souza et al. \cite{souza24casper} conduct centralized carbon forecasting, load balancing, and provisioning for distributed web services to minimize emissions while reducing the latency caused by load balancing. Kaur et al. \cite{kaur20keids-rt} facilitate real-time services across edge nodes integrating on-site renewable energy. In that, a centralized controller runs scheduling logic to schedule applications to meet energy and performance obligations.

Recent cloud-native and 6G frameworks employ centralized control planes to enforce global policy constraints and predictive orchestration. 
Carbon-Kube \cite{11434192} and CARC \cite{11542095} implement centralized Kubernetes controllers utilizing Custom Resource Definitions (CRDs) to enforce multi-objective DAG scheduling and SLO-aware carbon policies. 
Similarly, Sun et al. \cite{11619171} and Jiang et al. \cite{11666925} deploy centralized reinforcement learning agents to manage serverless pod retention and mobile IoT offloading. 
At a network scale, Batool et al. \cite{batool2026ai} utilize a logically centralized 6G control plane to coordinate multi-scale predictive scheduling across CPN tiers.

\noindent\textbf{Decentralized:}
In decentralized control, cloud environments achieve improved scalability and autonomy of individual elements. Further, it eliminates single points of failure in the system design. Nevertheless, decentralized control can introduce additional synchronization overheads between individual system elements. As a result, decentralized control is rarely applied in the literature. However, given that the carbon efficiency of renewable energy sources improves as they spread spatially, decentralized control can deliver scalable cloud environments that provide low-latency application performance while effectively utilizing geographically spread low-carbon intensity energy sources.

Sajid et al. \cite{sajid2021blockchain} propose a blockchain-based decentralized workload distribution and management model for geo-distributed data centers. It optimizes variability in renewable energy generation as a cost optimization problem and uses a blockchain model to employ a decentralized control architecture. Their work is designed for low-latency networks. Nevertheless, their approach can increase both request scheduling times and processing overhead, affecting application latency performance. To mitigate this, they design their management technique to optimize for both. Li et al. \cite{li2013chameleon} integrate renewable energy in a decentralized manner by allocating individual power budgets to servers and equipping each with a power management microcontroller. In return, each server aims to maximize renewable energy usage individually. Server performance, such as workload latency, is maintained through the power profiles of the microcontroller, which switches between an energy-oriented profile and a performance profile to maintain adequate workload performance while utilizing both renewable energy and battery-stored energy. Agarwal et al. \cite{agarwal23hvm} similarly integrate renewable energy. They drive the availability of CPU cores to match renewable energy availability individually at the server level in a decentralized manner. They modify the virtualization layer to efficiently manage dynamic CPU core availability so that virtual machines can continue executing, eliminating service interruptions.

Decentralized control is increasingly adopted to strictly respect multi-domain trust boundaries and overcome scale barriers. 
Al-Shammaa and Al-Hadrawi \cite{ALSHAMMAA2026101398} design a decentralized federated orchestrator that coordinates workload placement across autonomous edge-cloud domains without violating data sovereignty or requiring centralized telemetry aggregation. 
Likewise, Gao et al. \cite{11407972} propose an online distributed scheduling mechanism based on Lyapunov drift-plus-penalty optimization to resolve local device and edge server subproblems independently per time slot.

\subsubsection{System Tier}

Carbon-aware resource management techniques can be applied at multiple system tiers in cloud environments. Given the diverse latency SLOs of latency-sensitive applications, each tier offers different optimization opportunities, often specific to application dynamics. 
In this section, we discuss relevant works in the literature and examine how their opportunities, constraints, and latency implications vary across system tiers.
\noindent\textbf{Server Level:}
Server-level integration of renewable energy commonly involves improving the power delivery architecture of the data center to supply a particular renewable energy allocation to each server. Then, server-level workload management ensures that the server load adheres to the energy allocation. As a result, power management can be conducted without migrating workloads between servers, eliminating associated temporary service blackouts that impact application latency performance.

Li et al. \cite{li2013chameleon} leverage server-level energy allocation and introduce a micro-controller in each server to match the energy dynamics by throttling the server performance when needed, given that the workload throughput constraints allow it. Otherwise, battery-stored energy is used to match energy availability. Similarly, Agarwal et al. \cite{agarwal23hvm} use a server-level renewable energy budget. They employ dynamic core availability for running workloads to throttle at the server level when needed. In return, workload migration and the associated interruptions are avoided.

\noindent\textbf{Data Center Level:}
Data center-level integration of renewable energy allows resource management techniques to absorb the variable availability of renewable energy by adjusting the power consumption of servers as a whole. In return, resource management techniques can explore utilization patterns across the data center servers to adjust their power consumption while minimizing application latency performance.

Sharma et al. \cite{sharma11blink} employ an application-independent power management framework that allows defining policies to engage servers in the data center to yield a duty cycle of activation and deactivation, thus reducing the average power draw of the data center to match renewable energy availability. They demonstrate that specific distributed latency-sensitive applications integrated into the framework through their APIs can adapt their application patterns to deliver acceptable performance. Govindan et al. \cite{govindan11stored-energy-dc} apply energy buffers in the data center power delivery with UPS batteries to absorb renewable energy variations. They use server throttling at the data center level to optimize battery life and application performance, such as workload latency. Souza et al. \cite{souza23ecovisor} leverage data center-level renewable energy integration to provide virtualized energy systems to applications. In return, individual applications executing in the data center manage grid energy, renewable energy, and batteries through virtualization, employing application-specific carbon budgeting policies that allow them to meet their latency constraints.

At the data center and cluster level, recent containerized and serverless platforms integrate carbon telemetry into runtime management. 
Almaipeta et al. \cite{11542095} implement cluster-level microservice replica scaling and background job deferral within enterprise Kubernetes environments. 
Complementarily, Sun et al. \cite{11619171} optimize pod-level keep-alive durations within serverless data center clusters to minimize idle baseload carbon emissions.

\noindent\textbf{Data center and Server Levels:}
Given application characteristics, integrated management at both data center and server levels can yield better integration of renewable energy. In that context, server-level integration can enable workload management, avoiding workload migrations, while data center-level management can complement it across the servers.

Jahanshahi et al. \cite{jahanshahi2022powermorph} design a power reshaping framework for data centers catering to energy capacity signals from the power grid, such as the varying capacity of renewable energy integration. They execute both latency-sensitive and complementary workloads on servers. Due to short intervals of power shaping and long intervals of power bidding with the grid, they employ server-level power management via workload throttling and core allocation among latency-sensitive and complementary workloads while conducting data center-level power bidding at longer intervals.

\noindent\textbf{Inter-cloud Level:}
Integrating renewable energy at the inter-cloud level allows renewable energy availability to be utilized in different geographical locations. Here, the challenging aspect is maintaining the application performance of latency-sensitive workloads over the communication overhead of the network that connects geographical locations. In this context, resource management techniques aim to tackle the variable availability of renewable energy by directing workloads to locations with sufficient energy availability while adhering to latency constraints.

Yuan et al. \cite{yuan2019delaytolerantgeodatacenters} conduct spatiotemporal task scheduling across geographically distributed green data centers. Their task scheduling problem applies a delay-bound constraint to maintain workload latency performance, and scheduling algorithms are designed to adhere to that. Sajid et al. \cite{sajid2021blockchain} design their work for high-performance inter-cloud networks. Hence, they optimize application latency performance by minimizing both processing overhead and inter-cloud workload migration delays. Sun et al. \cite{sun24modular-datacenters} co-locate complementary workloads with latency-sensitive workloads in geo-distributed clouds and aim to reduce the workload migration frequency of latency-sensitive workloads to minimize their latency impact. Sukprasert et al. \cite{sukprasert24limits-of-load-shift} explore inter-cloud workload execution for latency-sensitive interactive workloads to harness renewable energy. They migrate interactive workloads to greener locations if latency constraints allow it. Murillo et al. \cite{murillo24cdnshifter} apply inter-cloud renewable energy harnessing to content delivery networks. They combine request latency-aware workload shifting with VM capacity shifting while maximizing renewable energy utilization within latency boundaries. Gsteiger et al. \cite{gsteiger24caribou} utilize inter-cloud renewable energy availability for serverless applications. They segregate functions from serverless application workflows that can be offloaded without increasing end-to-end latency and use that to utilize energy availability across clouds. Souza et al. \cite{souza24casper} harness renewable energy across clouds for distributed web services. They provision resources based on carbon forecasting and conduct load balancing across clouds within the latency constraints. Chien et al. \cite{chien23genai-rq-direct} investigate inter-cloud serving of generative AI requests. Directing requests to locations with minimum carbon intensity shows that renewable energy can be utilized without significantly impacting request latency. Kaur et al. \cite{kaur20keids-rt} leverage a multi-cluster deployment across different locations to execute real-time services. They propose a controller to utilize renewable energy across locations while maintaining adequate application performance.

Operational carbon management has expanded beyond traditional inter-cloud data centers toward the Cloud-Edge-Device continuum and 6G network fabrics. 
Al-Shammaa and Al-Hadrawi \cite{ALSHAMMAA2026101398} and Bhat et al. \cite{11434192} optimize workload placement across geo-distributed cloud regions and federated edge domains. 
Jiang et al. \cite{11666925} and Gao et al. \cite{11407972} target the Cloud-Edge-Device continuum, orchestrating offloading across mobile IoT devices, multi-edge nodes, and central cloud data centers. 
Finally, Batool et al. \cite{batool2026ai} unify computing power and network slicing over 6G CPNs to manage ultra-low-latency inter-tier execution.

\subsubsection{Latency Tolerance}

Application workloads in latency-sensitive cloud environments exhibit various SLOs regarding their latency performance. We classify these workloads into two primary categories: low-latency and bounded-latency. Low-latency applications prioritize short response times but can tolerate limited degradation within application-specific margins. For these applications resource management techniques seek to minimize such degradation and improve service quality. In contrast, bounded-latency applications impose response-time limits that must be strictly respected, meaning that carbon-aware decisions cannot introduce delays that would violate their SLOs. For these applications, including real-time workloads, resource management techniques must guarantee that response-time bounds are met. In this section, we examine how existing approaches exploit the variable availability of renewable energy while satisfying the latency requirements of both workload categories.

\noindent\textbf{Low-latency:}
Low latency applications that tolerate certain levels of latency degradation provide better flexibility in adapting to performance optimization of variable-available renewable energy integrations. As a result, most related works exploit this in two primary aspects: server performance throttling to match renewable energy availability and workload migration across geographical locations for greener energy availability.

Li et al. \cite{li2013chameleon} leverage dynamic throttling of server performance to engage an energy-efficient power profile that harnesses renewable energy. They employ techniques to dynamically switch between a high-performance power profile to minimize the impact of workload latency from server performance throttling. Jahanshahi et al. \cite{jahanshahi2022powermorph} use Dynamic Voltage Frequency Scaling (DVFS) to throttle CPU core performance, matching renewable energy availability. They minimize latency increases for latency-critical workloads by throttling cores allocated to complementary workloads first. Similarly, Govindan et al. \cite{govindan11stored-energy-dc} use DVFS-based throttling to optimize between latency performance and the longevity of battery energy storage, and Agarwal et al. \cite{agarwal23hvm} use throttling through shrinking CPU core availability to exploit latency performance for renewable energy harnessing. Sharma et al. \cite{sharma11blink} use application-independent blinking of server activation to match renewable energy availability and offload optimizing latency performance over throttling to the application. A similar approach can be seen with Souza et al. \cite{souza23ecovisor}, where a virtualized energy system allows applications to manage the throttling of their containers according to workload patterns while minimizing the latency performance impact over variable availability of renewable energy. 

Besides server performance throttling, many works exploit low latency for renewable harnessing across geographical locations. Sajid et al. \cite{sajid2021blockchain} exploit latency flexibilities in workloads with intermittent interruptions in workload availability to shift and execute them in greener data centers. They propose an optimized scheduling approach to minimize the interruption durations. A similar workload migration approach is used by Sun et al. \cite{sun24modular-datacenters} for modular data centers. They prioritize migrating VMs with fewer VM states to reduce the impact of migration overhead on latency performance. Murillo et al. \cite{murillo24cdnshifter} explore adjusting response latency time in content delivery networks to better harness renewable energy across edge data centers. They show that increasing latency within safe limits can reduce carbon emissions by up to 35.5\%. Gsteiger et al. \cite{gsteiger24caribou} leverage low latency in serverless applications for carbon optimization by segregating functions in application workflows and offloading those with lesser latency impact to greener locations. Souza et al. \cite{souza24casper} increase request latency in distributed web services for renewable energy harvesting across locations by combining load balancing and resource provisioning. Chien et al. \cite{chien23genai-rq-direct} apply renewable energy harnessing across geo-spread locations to serve generative AI requests and show that the low-latency characteristics of requests allow for operation within safe limits.
Almaipeta et al. \cite{11542095} leverage low SLO risk windows to downscale microservice replicas and defer background tasks, while Sun et al. \cite{11619171} dynamically tune serverless keep-alive times to accept controlled cold-start latencies during high carbon intensity periods. 
In mobile IoT systems, Jiang et al. \cite{11666925} and Gao et al. \cite{11407972} maintain stable task queue backlogs while offloading latency-flexible tasks across multi-edge and device tiers.

\noindent\textbf{Bounded-latency:}
Exploiting applications with bounded latency constraints for renewable energy integration is less common. This is due to the rigid nature of latency upper bounds, which do not provide much room for resource management techniques to absorb renewable energy variations.

Yuan et al. \cite{yuan2019delaytolerantgeodatacenters} leverage a scheduling algorithm that strictly guarantees the task's delay-bound constraints while maximizing the usage of renewable energy across distributed green data centers. Kaur et al. \cite{kaur20keids-rt} explore real-time services with well-defined response time boundaries for harvesting renewable energy across Kubernetes clusters with on-site renewable energy integration. They propose a controller to optimize green energy utilization and performance impacts from application inferences in container scheduling.
Batool et al. \cite{batool2026ai} guarantee sub-millisecond real-time response bounds for 6G Ultra-Reliable Low Latency Communications safety-critical applications via proactive AI prediction. 
Bhat et al. \cite{11434192} enforce rigid completion SLAs for big data Spark DAG pipelines using forecast-aware temporal planning without violating execution envelopes. 
Furthermore, Al-Shammaa and Al-Hadrawi \cite{ALSHAMMAA2026101398} implement an emergency mode with zero SLA violations for mission-critical federated workloads.

\subsection{Embodied Carbon Management}

The management of embodied carbon is broad, spanning across the lifecycle of IT assets. We classify it into two primary stages: pre-deployment and post-deployment. We then discuss related resource management techniques related to planning, designing, managing, and recycling IT assets in cloud environments. We refer to the primary stages that we identified while narrowing our focus to those considering the impact on application latency performance.

\subsubsection{Pre-deployment}

Since embodied carbon management is conducted through the management of IT assets, the pre-deployment stage of cloud environments approaches it at the infrastructure design stage. A well-designed infrastructure for carbon efficiency reduces the need to optimize again once the infrastructure is available to execute workloads. Here, we focus on two primary aspects: infrastructure design and low-carbon component use, emphasizing the latency performance of running workloads as a design goal.

\noindent\textbf{Infrastructure Design:}
Cloud environments can exhibit heterogeneous designs depending on the workload performance requirements. Each design can offer specific opportunities to improve embodied carbon efficiency. Moreover, specific tools for estimating embodied carbon footprints can lead to better designs. Finally, techniques that maintain adequate application latency performance with a reduced hardware footprint can also reduce the embodied carbon footprint.

Sun et al. \cite{sun24modular-datacenters} investigate geographically distributed data center networks capable of serving low-latency workloads. They use modular data centers that integrate renewable energy sources (rMDC) in their design. rMDC incurs a lesser embodied carbon footprint than traditional data centers. Their work enables colocating rMDC with more stable energy sources, reducing the number of servers required to utilize peak energy spikes, thus further reducing the embodied carbon. Ji et al. \cite{ji24scarif-embodied-design} design a carbon estimation tool for servers with accelerators, such as GPU and FPGA, that serves latency-sensitive workloads. They advance accurate embodied carbon estimation and enable operators to derive improved carbon efficiency in their designs. Tannu et al. \cite{tannu23dirty-ssd} explore embodied carbon intensity in storage mediums. Their work provides insightful guidance on selecting carbon-efficient storage mediums like HDD or SSD. SSDs provide better performance for workload latency than HDDs, yet their embodied carbon efficiency can differ. Therefore, carbon insights help operators make design decisions to balance workload performance and embodied carbon efficiency. Gupta et al. \cite{gupta24meta-embodied} aim to shrink the overall hardware footprint of the cloud environment using resource oversubscription, reducing underutilization and embodied carbon footprint. They design a dynamic system to observe resource underutilization and present a dynamic resource leasing platform. Through that, their system can execute latency-sensitive workloads with capacity availability SLOs over the oversubscribed infrastructure. Results show that infrastructure shrinking by as much as 25\% is achievable.

\noindent\textbf{Low-carbon Component Use:}
Opting for components with lower embodied carbon footprints enables operators to optimize the carbon efficiency of their cloud environment design. However, the associated performance bottlenecks must be identified, and suitable techniques must be employed to mitigate them.

Zhong et al. \cite{yuhong24cxl-memory} propose a tiered memory system that reduces the embodied carbon footprint. Instead of local DRAM, they use a hardware-managed tiered memory system for low-carbon CXL-based memory. Although CXL has a lower embodied carbon footprint, it can result in higher latencies than local DRAM. They introduced a software stack to manage this issue. The combination of the hardware-managed tiering system and the software stack provides memory performance that is closer to that of local DRAM.

\subsubsection{Post-deployment}

Once deployed, a cloud environment's embodied carbon footprint can be optimized by managing installed IT assets. Most works aim to extend the asset's operating life or reuse older components. At the post-deployment stage, the embodied carbon footprint has already been established. Thus, extending the asset's lifetime or reusing older components allows operators to amortize the acquired carbon further. However, using aged components can increase the risk of component failure in the cloud environment. Resource management techniques must mitigate this risk by exploring opportunities in hardware-software settings while maintaining adequate workload latency performance.

\noindent\textbf{Lifetime Extension:}
Resource management techniques that optimize the embodied carbon footprint through the asset's lifetime extension exploit resource usage patterns specific to the hardware in focus. They manipulate resource usage patterns, and through that, they improve asset longevity. The resulting lifetime extension allows for further amortization of the asset's embodied carbon footprint.

Zhao et al. \cite{zhao2023unsustainableaffinity} aim to reduce uneven core wear-off in multi-core CPUs from executing workloads with core affinity, such as real-time workloads requiring deterministic performance. In return, premature failures of specific cores can be avoided for extended CPU life. They provide a performance metric independent of CPU micro-architectural characteristics to measure uneven CPU core usage, which can be used at the cloud resource management layer. Leveraging the proposed metric, the workload scheduler can shift workloads between CPU cores based on CPU stress. Wang et al. \cite{wang23peeling-carbon} employ an aging-aware workload scheduler to maintain workload performance among servers with heterogeneous aging characteristics. They identify older servers that could maintain sufficient performance under specific conditions, such as during low load times, and use that knowledge in the workload scheduler. Tannu et al. \cite{tannu23dirty-ssd} reduce SSD wear-off in storage server fleets. They employ dynamic data redirection to locations with lower write intensity. This evens out SSD writing across the server fleet, reducing their aging rate, and allowing for an extended amortization of embodied carbon. Similarly, Gupta et al. \cite{gupta22act} reduce the SSD aging rate by increasing over-provisioning in their cloud environment setting to reduce the write amplification factor. McAllister et al. \cite{sara24wrencache} exploit premature failure risks in flash-based cache. Instead of a legacy logical block addressable device interface, they propose fairyWREN, a flash cache designed for write read erase interface (WREN). WREN allows application control over data placement and garbage collection, which fairyWREN uses to reduce writes via caching policies. fairyWREN has better read latency at peak load, improving the latency-sensitive performance of workloads and extending flash storage lifetime for improved embodied carbon amortization. 

\noindent\textbf{Component Reuse:}
Used components can be employed in data centers to reduce the acquired embodied carbon of newer components, given that the workload performance, such as latency requirements, is maintained. Used components have already amortized their initial embodied carbon footprint to a certain degree; thus, using them reduces the overall embodied carbon footprint of the cloud environment.

Wang et al. \cite{wang24greensku} propose a framework to evaluate carbon savings at scale with used components, such as used memory and SSD. It enables providers to evaluate performance at scale with used components, such as tail latency and low load latency. In turn, operators can make informed decisions about using these used components in their cloud environments. Tannu et al. \cite{tannu23dirty-ssd} propose re-purposing used flash devices. They focus on Multi-level cell (MLC) devices, which store multiple bits within each cell, thus providing higher capacities. However, MLC can wear out rapidly compared to single-level cell devices (SLC). The work proposes a strategy to transform used MLC into low-capacity SLC devices, enabling a second life to amortize embodied carbon. Chien et al. \cite{chien23genai-rq-direct}  explore embodied carbon optimization for the geo-distributed serving of low-latency generative AI requests, where used servers provide headroom in each location. The work evaluates the amount of headroom needed for effective carbon improvements. Gupta et al. \cite{gupta22act} explore reusing general-purpose hardware instead of employing specialized accelerators. Their work shows a promising balance between component reuse and workload performance.

\section{Classification of Resource Management Techniques Using the Proposed Taxonomy}
\label{sec:classification-of-rm-techniques}

Table \ref{tab:rm-classify-op-carbon} and \ref{tab:rm-classify-em-carbon} outline key works on carbon-aware resource management in latency-sensitive cloud environments related to the proposed taxonomy. Table \ref{tab:rm-classify-op-carbon} focuses on operational carbon management, and Table \ref{tab:rm-classify-em-carbon} focuses on embodied carbon management. The works we present here propose novel resource management techniques that explore one or more of the resource management aspects we have identified. 

\begin{landscape}
    \scriptsize
    \begin{longtable}
                {>{\raggedright\arraybackslash}p{0.04\linewidth}|>{\centering\arraybackslash}p{0.105\linewidth}|>{\centering\arraybackslash}p{0.075\linewidth}|>{\centering\arraybackslash}p{0.060\linewidth}|>{\centering\arraybackslash}p{0.09\linewidth}|>{\centering\arraybackslash}p{0.095\linewidth}|>{\centering\arraybackslash}p{0.075\linewidth}|>{\centering\arraybackslash}p{0.095\linewidth}|>{\centering\arraybackslash}p{0.065\linewidth}|>{\centering\arraybackslash}p{0.075\linewidth}} 
                \hline 
                Work&  \multicolumn{3}{|c|}{Workload Management}&  Control & \multicolumn{3}{|c|}{Environment}&\multicolumn{2}{c}{Latency SLO}\\ 
                \hline
                &  Load Matching & Granularity & Mixed Criticality & & System Tier & Topology & Kind & Tolerance & Optimize \\
                \hline
                \cite{sun24modular-datacenters}& Load Shifting, Preemption& VM& Yes& Centralized& Inter-Cloud& Networked Clouds& Modular DCs& Low-latency& Downtime, Criticality\\

                \cite{murillo24cdnshifter}& Load Shifting & Server Load & - & Centralized& Inter-Cloud& Networked Clouds & CDNs & Low-latency & Response Latency \\

                \cite{gsteiger24caribou}& Load Shifting & Function & Yes & Centralized & Inter-Cloud & Networked Clouds & Serverless & Low-latency & End-to-end Latency \\

                \cite{souza24casper}& Load Shifting & Request & - & Centralized & Inter-Cloud & Networked Clouds & Distributed Web Services & Low-latency & Response Latency \\

                \cite{souza23ecovisor}& Throttling & VM, Container & - & Centralized & Data Center-level, Server-level & Server Fleets & Virtualized or Containerized Apps  & Low-latency & Response Latency \\

                \cite{chien23genai-rq-direct}& Admission Control and Placement & Request & - & - & Inter-Cloud & Networked Clouds & Gen. AI Serving & Low-latency & Response Latency \\

                \cite{jahanshahi2022powermorph}& Throttling & Task & Yes & Centralized & Data Center-level, Server-level & Server Fleets & - & Low-latency & Response Latency \\

                \cite{sajid2021blockchain}& Load Shifting & Request & - & Decentralized & Inter-Cloud & Networked Clouds & - & Low-latency & Migration Latency \\

                \cite{kaur20keids-rt}& Admission Control and Placement & Container & - & Centralized & Inter-Cloud & Networked Clouds & Real-Time Services & Bounded-latency & Response Latency \\

                \cite{yuan2019delaytolerantgeodatacenters}& Admission Control and Placement & Task & - & Centralized & Inter-Cloud & Networked Clouds & Green DCs & Bounded-latency & Response Latency \\

                \cite{li2013chameleon}& Throttling & Server Load & Yes & Decentralized & Server-level & Server Fleets & Throughput Servers & Low-latency & Response Latency \\

                \cite{sharma11blink}& Throttling & Server Load & - & Centralized &Data Center-level & Server Fleets & Interrupt-tolerable Apps & Low-latency & Response Latency \\

                \cite{govindan11stored-energy-dc}& Throttling & Server Load & - & Centralized &Data Center-level & Server Fleets & - & Low-latency & Response Latency \\

                \cite{agarwal23hvm}& Throttling & VM & - & Decentralized & Server-level & Server Fleets & Virtualized Apps & Low-latency & Response Latency \\

                \cite{batool2026ai}& Load Shifting & CPN Task / Slice & Yes & Centralized & Inter-Cloud & Networked CPN & 6G CPN & Bounded-latency & Proactive Latency, Mobility \\

                \cite{11666925}& Admission Control and Placement & Task & Yes & Centralized & Inter-Cloud & Cloud-Edge Continuum & Mobile IoT & Low-latency & Long-term Carbon Budget \\

                \cite{11407972}& Admission Control and Placement & Task & - & Decentralized & Inter-Cloud & Cloud-Edge-Device & IoT Systems & Low-latency & Green Energy Harvesting \\

                \cite{11542095}& Throttling & Microservice Replica & Yes & Centralized & Data Center-level & Server Fleets & Enterprise Microservices & Low-latency & SLO Risk, Carbon Cost \\

                \cite{11434192}& Load Shifting & Pipeline DAG & Yes & Centralized & Inter-Cloud & Networked Clouds & Big Data Pipelines & Bounded-latency & Data Gravity, SLA \\

                \cite{ALSHAMMAA2026101398}& Admission Control and Placement & Workload / Task & Yes & Decentralized & Inter-Cloud & Federated Edge-Cloud & Multi-Domain Edge-Cloud & Bounded-latency & Multi-Objective, Sovereignty \\

                \cite{11619171}& Throttling & Function Pod & Yes & Centralized & Data Center-level & Server Fleets & Serverless / FaaS & Low-latency & Cold Start vs Idle Carbon \\
                
                \hline
        \\
        \caption{Classification of Resource Management Techniques of Operational Carbon Reduction}

        \label{tab:rm-classify-op-carbon}
    \end{longtable}
\begin{longtable}{%
    >{\raggedright\arraybackslash}p{0.04\linewidth}|
    >{\centering\arraybackslash}p{0.11\linewidth}|
    >{\centering\arraybackslash}p{0.11\linewidth}|
    >{\centering\arraybackslash}p{0.11\linewidth}|
    >{\centering\arraybackslash}p{0.11\linewidth}|
    >{\centering\arraybackslash}p{0.11\linewidth}|
    >{\centering\arraybackslash}p{0.11\linewidth}|
    >{\centering\arraybackslash}p{0.11\linewidth}
} 
    \hline
    Work & Approach & \multicolumn{2}{|c|}{Pre-deployment} & 
    \multicolumn{2}{|c|}{Post-deployment} & 
    \multicolumn{2}{|c}{Optimize} \\ 
    \hline
    & & Infrastructure Design & Low-carbon Components & Extended Life & Component Reuse & Hardware Aspect & App Latency \\ 
    \hline
    \cite{sun24modular-datacenters} & Designing, Planning & Energy Stability & Modular DC & - & - & Datacenter Size & Network Delay \\ 
    \cite{ji24scarif-embodied-design} & Carbon Estimation & Carbon vs Performance & - & - & - & Accelerators & Accelerated Computing \\ 
    \cite{gupta24meta-embodied} & Resource Oversubscription & Size Reduction & - & - & - & Utilization & Capacity-availability SLOs \\ 
    \cite{tannu23dirty-ssd} & Component Selection, Workload Scheduling & Carbon vs Performance & - & SSD Aging & Re-purpose Flash Devices & Storage Medium, Write Intensity & Storage Access \\ 
    \cite{yuhong24cxl-memory} & Component Selection & - & CXL Memory & - & - & Memory & Memory Access \\ 
    \cite{zhao2023unsustainableaffinity} & Workload Scheduling & - & - & CPU Aging & - & CPU Wear-off & Scheduling Overhead \\ 
    \cite{wang23peeling-carbon} & Workload Scheduling & - & - & Server Aging & - & Servers & Age vs Performance \\ 
    \cite{gupta22act} & Resource Over provisioning & - & - & SSD Aging & General-purpose H/W for Accelerators & Storage Write Intensity, Accelerators & Storage Access, General-purpose H/W vs Performance \\ 
    \cite{sara24wrencache} & Workload Scheduling & - & - & Flash Device Aging & - & Storage Write Intensity & Storage Access \\ 
    \cite{wang24greensku} & Carbon Estimation & - & - & - & Used Components & Reusing at Scale & Age vs Performance \\ 
    \cite{chien23genai-rq-direct} & Workload Scheduling & - & - & - & Used Servers & Data Center Headroom & Age vs Performance \\ 
    \hline
    \\
    \caption{Classification of Resource Management Techniques for Embodied Carbon Reduction}
    \label{tab:rm-classify-em-carbon}
\end{longtable}

\end{landscape}

\section{Research Gaps and Future Directions}
\label{sec:research-gaps}

The in-depth review of carbon-aware resource management in latency-sensitive cloud environments that we conducted highlights open problems with great potential for further exploration. This section provides an in-depth discussion of these areas and the broader categories identified in our taxonomy, considering both operational and embodied carbon efficiency. Through this analysis, we lay the groundwork for future research and development.

\subsection{Spatio-Temporal Matching of Intermittent Clean Energy Across Continuous Tiers}
\par Emerging low-latency applications, such as the Internet of Things (IoT), are executed in cloud deployments that rent space and power in distributed, resource-constrained environments, such as colocation data centers. As tenants, clouds subscribe to limited power budgets from the colocation provider, which may integrate on-site renewable energy into its shared power delivery to improve carbon efficiency. 
As a result, tenant clouds face the challenge of optimizing the use of renewable energy through shared power delivery. This optimization must be achieved not only locally but also across distributed deployments of similar clouds, which provide varying levels of renewable energy availability depending on time and location.

\par Most low-carbon intensive renewable energy integration solutions for cloud environments use data center-level allocation of renewable energy. 
Although this allows for workload management  across servers or even between clouds to align with renewable energy capacity fluctuations, renewable energy-aware allocation strategies at server-level or core-level can benefit applications with strict latency SLOs by preventing workload migrations. 
However, uneven workload allocation across servers due to dynamic resource usage in clouds may result in underutilization of renewable energy at the server or core level. Improving this through resource management is an interesting open problem.

\par 
Furthermore, the integration of dynamic, multi-source energy harvesting (e.g., on-site solar and wind) into heterogeneous edge-cloud environments introduces severe stochasticity. 
While edge servers leveraging green energy harvesting can significantly reduce reliance on carbon-intensive grids, matching intermittent renewable availability with multi-tier queue stability across Cloud-Edge-Device continuum remains an open challenge.
Future research must address how multi-domain infrastructures can balance localized clean energy availability against spatial grid carbon variations without over-provisioning energy storage or incurring excessive cooling overheads.

\subsection{Mitigating Embodied Carbon Through Hardware Longevity and Thermal Load Management}

\par Increasingly popular generative AI-based applications demand low-latency responses, which require cloud servers with accelerators for faster model computation. Although existing works have explored better energy efficiency through older servers by segregating generative AI model computation workflows, the embodied carbon impact has not been thoroughly examined. For instance, defining latency SLOs for generative AI model inference requests can better inform the request scheduler to balance performance and carbon efficiency among older and newer generation servers. This enables cloud environments to deploy older generation servers or used servers to reduce their embodied carbon footprint.

\par 
Complementing hardware reuse strategies, managing the physical degradation induced by aggressive power configurations represents a critical pathway to mitigating embodied carbon. Meeting rigid latency SLOs often forces providers to tune servers and accelerators to high-performance, peak-power profiles, exposing underlying components to sustained thermal and electrical stress over extended periods. As a result, this accelerated wear causes thermal hotspots and premature component failures, forcing cloud providers into frequent hardware replacements that substantially inflate the environment's net embodied carbon footprint. For example, the uneven distribution of CPU stress from infrastructure tasks, such as virtualization and workload tasks, has been identified as a cause of thermal hotspots in CPU cores. In this context, further exploring stress-aware resource management techniques that balance high-performance SLO compliance with component longevity becomes paramount to curbing embodied carbon emissions.

\par Opportunities for reducing embodied carbon through server longevity can be seen in the growing prevalence of generative AI inference clusters, which offload most computational stress to accelerators such as GPUs. As a result, resource usage patterns in such clusters can lead to the underutilization of resources such as CPUs. Improving resource management techniques to identify underutilization patterns and leverage them to reduce computational stress in server components can improve their longevity, leading to amortizing embodied carbon over the improved lifetime.

\subsection{Harnessing Workload Criticality and Fault Tolerance for Carbon Efficiency}

\par 
Cloud providers increasingly offer infrastructure-as-a-service (IaaS) mechanisms that expose application criticality tiers to the cloud management plane, such as evictable or spot virtual machines. Workloads in enterprise clouds span diverse criticality levels, ranging from mission-critical transactional services that demand uninterrupted execution to fault-tolerant, non-critical tasks such as background analytics, log processing, or model re-training. Leveraging these criticality distinctions alongside application-level fault tolerance to absorb renewable energy intermittency remains a major open challenge. For example, application-specific middleware can deploy latency-sensitive workflows across evictable VMs, dynamically shedding or deferring low-criticality instances when clean energy drops while shielding high-priority components. In this context, exploiting application-specific SLOs to integrate renewable energy with VM evictability as a load-matching mechanism requires novel resource management techniques.

\par 
Beyond evictable VMs, exploring how heterogeneous application criticality classes absorb renewable energy variations without degrading user experience remains a key open direction. While many renewable energy integration techniques focus on flexible, low-latency applications that can easily tolerate minor delays, workloads with bounded latency SLOs (such as real-time industrial automation, interactive streaming, or safety-critical edge services) correspond to a significant portion of cloud computing's operational carbon footprint. Exploring potential opportunities to absorb the variable availability of renewable energy sources across bounded-latency applications, by dynamically adjusting redundancy, execution modes, or performance profiles without violating rigid response boundaries, becomes paramount.

\par 
In addition to technical scheduling safeguards, widespread adoption of carbon-aware cloud services requires robust economic incentive mechanisms that encourage end-user participation. 
Cloud tenants often resist carbon-driven execution delays or dynamic SLO relaxation without tangible compensation. 
Implementing dynamic carbon-pricing tiers, green SLA contracts, and carbon credits, where users receive billing discounts, priority credits, or tax offsets in exchange for accepting flexible completion windows during high-carbon grid hours, represents a vital interdisciplinary research direction. 
Designing game-theoretic incentive structures that align tenant economic objectives with provider decarbonization targets will be essential for real-world production deployment.

\subsection{Mitigating Data Gravity and Cascading Failures in Dependent Workloads}

\par 
In complex multi-component cloud and edge architectures, latency SLOs are heavily dictated by structural and data dependencies across interconnected services. 
Spatial load shifting is frequently constrained by Data Gravity, the prohibitive network latency, transfer overhead, and bandwidth costs of relocating large data volumes or stateful dependencies across geo-distributed regions. 
Co-optimizing carbon-aware temporal and spatial planning while explicitly accounting for data gravity penalties and strict end-to-end SLAs remains a fundamental research gap. 
Furthermore, in graph-structured workflows and microservice call chains, applying localized carbon-aware actions (such as aggressive downscaling or throttling) on upstream components risks retry amplification and cascading SLO failures across downstream dependencies. 
Future resource management frameworks must move beyond localized decisions to implement topology-aware safeguards that evaluate end-to-end application dependencies before applying carbon mitigation strategies.

\subsection{Balancing Responsiveness and Idle Footprints in Granular Serverless Runtimes}
\par Computing models such as serverless computing allow the resource management layer to make granular management decisions. For instance, instead of application-level scheduling, the resource management layer can schedule functions that may collectively conduct an application workflow. As a result, additional opportunities are available to make granular changes to the server power draw, which can match intermittent renewable energy availability. Exploiting that for various renewable energy allocation methods, such as server-level or core-level allocations, and various latency SLOs is an interesting research problem.

\par 
Granular computing paradigms, such as serverless FaaS, introduce a fundamental conflict between responsiveness and sustainability. 
To eliminate sub-second cold-start latency, platforms maintain warm function instances in memory after invocation completion. However, prolonged retention of idle execution environments incurs a substantial hidden carbon footprint due to continuous baseline power draw. 
Because function reuse intervals, runtime overheads, and cold-start penalties vary drastically across serverless workflows, static keep-alive timeouts fail to adapt to time-varying grid carbon intensity. 
Formulating adaptive runtime retention as a sequential decision problem that dynamically balances cold-start latency penalties against idle carbon emissions represents a vital research avenue for sustainable serverless platforms.

\subsection{Scaling Decentralized Orchestration and Preserving Multi-Domain Data Sovereignty}
\par Due to low-latency performance in executing applications at the network's edge, geographically distributed cloud environments are becoming increasingly popular. Nevertheless, carbon optimization in those deployments today mostly leverages centralized control, mainly for the ease of maintenance and management, yielding bottlenecks such as limited scaling capabilities. As a solution, decentralized control can be implemented. However, decentralized resource management requires efficient synchronization between clouds with minimum latency overhead. Exploring application-specific opportunities, such as executing with mixed latency SLOs of both low-latency and bounded components, can lead to better implementations of decentralized control.

\par As edge-cloud environments expand across organizational and geographical boundaries, carbon-aware orchestration inevitably crosses multi-domain federation trust boundaries. 
In these settings, traditional centralized orchestrators fail because they require aggregating global telemetry, which can violate data sovereignty laws (e.g., GDPR) and corporate privacy policies.
Future research must explore federation-aware placement algorithms that strictly enforce data sovereignty and domain-allowed constraints while optimizing spatio-temporal workload shifting. 
Developing lightweight, privacy-preserving coordination mechanisms that operate over multi-domain federations without incurring heavy consensus overheads or requiring raw data sharing remains an urgent open direction. 

\subsection{Ethical Considerations, Carbon Equity, and Global Social Impact}
\par As carbon-aware resource management scales across global cloud and edge deployments, spatio-temporal workload shifting also introduces critical ethical and socio-economic challenges. 
Relocating computationally intensive tasks across international borders to take advantage of clean energy resources in different regions can result in carbon leakage. This can displace operational footprints to developing regions with less stringent environmental regulations or put unexpected thermal and capacity stress on local power grids.
Furthermore, principles of global carbon equity dictate that energy-constrained or rural communities should not suffer degraded Quality of Service or reduced computing access due to aggressive carbon-capping policies enforced by multi-national cloud providers. 
Future resource management frameworks must incorporate fair-share carbon allocation models, regional grid-stability safeguards, and social equity metrics to guarantee that digital decarbonization does not exacerbate regional energy disparities or compromise local infrastructure resilience.

\subsection{Enforcing Long-Horizon Carbon Budgets in latency-sensitive scenarios}

\par While most existing schedulers minimize carbon footprint on a myopic, per-task basis, enterprise operators increasingly face long-term sustainability commitments, such as monthly carbon quotas or 24/7 Carbon-Free Energy (CFE) roadmaps. 
Bridging the gap between long-horizon carbon budgets and real-time latency-sensitive scheduling requires novel frameworks. 
However, integrating these long-term virtual budget constraints with multi-objective trade-offs (balancing carbon emissions, electricity costs, and strict SLA compliance) requires policy-as-code mechanisms and dynamic adaptive weighting that automatically adjust priorities under emergency or peak-carbon conditions.

\subsection{Ensuring Out-of-Distribution Robustness in AI-Native Predictive Management}

\par Shifting from reactive to anticipatory resource management is essential for ultra-low latency applications (such as 6G Computing Power Networks and autonomous IoT), where reactive allocation delays consume up to 80\% of the latency budget \cite{batool2026ai}. 
Multi-scale predictive frameworks combining short-term, spatial/mobility, and long-term strategic models have demonstrated great promise in positioning compute resources before demand peaks materialize.
However, a critical research gap lies in model robustness under cold-start conditions (e.g., lack of historical data for new users) and out-of-distribution (OOD) environmental shifts, such as sudden weather changes or solar droughts.
Future research should focus on confidence-based fallback mechanisms, circuit breakers, and online transfer learning to guarantee a deterministic performance floor when predictive AI models encounter anomalous grid or traffic behaviors.

\section{Summary}
\label{sec:summary}

In this article, we presented a detailed review of the aspect of carbon-aware resource management, focusing on latency-sensitive cloud environments. We proposed a taxonomy for a holistic view of carbon-aware resource management in both operational and embodied carbon efficiency. We discussed both aspects of operational and embodied carbon management and analyzed the existing works using the taxonomy. Our taxonomy presents an in-depth view of both operational and embodied carbon aspects for cloud operators to identify carbon optimization opportunities to meet their short-term and long-term carbon efficiency goals. Further, it provides the groundwork for researchers to understand existing works in carbon-aware resource management and to investigate and build upon their work. Finally, we provide a gap analysis highlighting the identified challenges, emphasizing the great potential for future work.

\begin{acks}
This work primarily funded by the Australian Research Council Discovery project. This work is also supported by the Spanish Ministry of Science, Innovation and Universities through the José Castillejo Mobility Programme for Young Researchers (grant CAS24/00117) for Patricia Arroba’s research stay at The University of Melbourne, and by the R\&D\&I Project PID2024-156033OA-C33, funded by MICIU/AEI/10.13039/501100011033 and by “ERDF/EU”.
\end{acks}

\bibliographystyle{ACM-Reference-Format}
\bibliography{refs}


\begin{thebibliography}{50}


\ifx \showCODEN    \undefined \def \showCODEN     #1{\unskip}     \fi
\ifx \showDOI      \undefined \def \showDOI       #1{#1}\fi
\ifx \showISBNx    \undefined \def \showISBNx     #1{\unskip}     \fi
\ifx \showISBNxiii \undefined \def \showISBNxiii  #1{\unskip}     \fi
\ifx \showISSN     \undefined \def \showISSN      #1{\unskip}     \fi
\ifx \showLCCN     \undefined \def \showLCCN      #1{\unskip}     \fi
\ifx \shownote     \undefined \def \shownote      #1{#1}          \fi
\ifx \showarticletitle \undefined \def \showarticletitle #1{#1}   \fi
\ifx \showURL      \undefined \def \showURL       {\relax}        \fi
\providecommand\bibfield[2]{#2}
\providecommand\bibinfo[2]{#2}
\providecommand\natexlab[1]{#1}
\providecommand\showeprint[2][]{arXiv:#2}

\bibitem[Agarwal et~al\mbox{.}(2023)]%
        {agarwal23hvm}
\bibfield{author}{\bibinfo{person}{Anup Agarwal}, \bibinfo{person}{Shadi
  Noghabi}, \bibinfo{person}{Inigo Goiri}, \bibinfo{person}{Srinivasan Seshan},
  {and} \bibinfo{person}{Anirudh Badam}.} \bibinfo{year}{2023}\natexlab{}.
\newblock \showarticletitle{Unlocking unallocated cloud capacity for long,
  uninterruptible workloads}. In \bibinfo{booktitle}{\emph{Proceedings of 20th
  USENIX Symposium on Networked Systems Design and Implementation (NSDI 23)}}.
  \bibinfo{publisher}{USENIX Association}, \bibinfo{address}{Boston, MA},
  \bibinfo{pages}{457--478}.
\newblock
\showISBNx{978-1-939133-33-5}
\urldef\tempurl%
\url{https://www.usenix.org/conference/nsdi23/presentation/agarwal-anup}
\showURL{%
\tempurl}


\bibitem[Al-Shammaa and Al-Hadrawi(2026)]%
        {ALSHAMMAA2026101398}
\bibfield{author}{\bibinfo{person}{Ameer~A. Al-Shammaa} {and}
  \bibinfo{person}{Doaa~H. Al-Hadrawi}.} \bibinfo{year}{2026}\natexlab{}.
\newblock \showarticletitle{Carbon-aware federated orchestration in edge-cloud
  systems: A multi-objective optimization framework with dynamic
  spatio-temporal workload management}.
\newblock \bibinfo{journal}{\emph{Sustainable Computing: Informatics and
  Systems}}  \bibinfo{volume}{51} (\bibinfo{year}{2026}),
  \bibinfo{pages}{101398}.
\newblock
\showISSN{2210-5379}
\urldef\tempurl%
\url{https://doi.org/10.1016/j.suscom.2026.101398}
\showDOI{\tempurl}


\bibitem[Almaipeta et~al\mbox{.}(2026)]%
        {11542095}
\bibfield{author}{\bibinfo{person}{Jagadish Almaipeta},
  \bibinfo{person}{Subhash Madineni}, {and} \bibinfo{person}{Usha~Rani Sheri}.}
  \bibinfo{year}{2026}\natexlab{}.
\newblock \showarticletitle{Carbon-Aware SLO Enforcement for Cloud-Native
  Microservices Using Adaptive Reliability Policies}. In
  \bibinfo{booktitle}{\emph{2026 International Conference on Artificial
  Intelligence, Systems, and Emerging Technologies (ICAISET)}}.
  \bibinfo{pages}{1--6}.
\newblock
\urldef\tempurl%
\url{https://doi.org/10.1109/ICAISET66439.2026.11542095}
\showDOI{\tempurl}


\bibitem[Ambati et~al\mbox{.}(2020)]%
        {ambati2020harvest-vms}
\bibfield{author}{\bibinfo{person}{Pradeep Ambati}, \bibinfo{person}{Íñigo
  Goiri}, \bibinfo{person}{Felipe Vieira~Frujeri}, \bibinfo{person}{Alper Gun},
  \bibinfo{person}{Ke Wang}, \bibinfo{person}{Brian Dolan},
  \bibinfo{person}{Brian Corell}, \bibinfo{person}{Sekhar Pasupuleti},
  \bibinfo{person}{Thomas Moscibroda}, \bibinfo{person}{Sameh Elnikety},
  \bibinfo{person}{Marcus Fontoura}, {and} \bibinfo{person}{Ricardo
  Bianchini}.} \bibinfo{year}{2020}\natexlab{}.
\newblock \showarticletitle{Providing SLOs for Resource-Harvesting VMs in Cloud
  Platforms}. In \bibinfo{booktitle}{\emph{Proceedings of the Symposium on
  Operating Systems Design and Implementation (OSDI)}}.
\newblock


\bibitem[Andreoli et~al\mbox{.}(2025)]%
        {andreoli2025multi}
\bibfield{author}{\bibinfo{person}{R. Andreoli}, \bibinfo{person}{R. Mini},
  \bibinfo{person}{P. Skarin}, \bibinfo{person}{H. Gustafsson},
  \bibinfo{person}{J. Harmatos}, \bibinfo{person}{L. Abeni}, {and}
  \bibinfo{person}{T. Cucinotta}.} \bibinfo{year}{2025}\natexlab{}.
\newblock \showarticletitle{A multi-domain survey on time-criticality in cloud
  computing}.
\newblock \bibinfo{journal}{\emph{IEEE Transactions on Services Computing}}
  \bibinfo{volume}{18}, \bibinfo{number}{2} (\bibinfo{year}{2025}),
  \bibinfo{pages}{1152--1170}.
\newblock
\urldef\tempurl%
\url{https://doi.org/10.1109/TSC.2025.3539197}
\showDOI{\tempurl}


\bibitem[Arroba et~al\mbox{.}(2024)]%
        {arroba2024sustainable}
\bibfield{author}{\bibinfo{person}{Patricia Arroba}, \bibinfo{person}{Rajkumar
  Buyya}, \bibinfo{person}{Román Cárdenas}, \bibinfo{person}{José~L.
  Risco-Martín}, {and} \bibinfo{person}{José~M. Moya}.}
  \bibinfo{year}{2024}\natexlab{}.
\newblock \showarticletitle{Sustainable edge computing: Challenges and future
  directions}.
\newblock \bibinfo{journal}{\emph{Software: Practice and Experience}}
  \bibinfo{volume}{54}, \bibinfo{number}{11} (\bibinfo{year}{2024}),
  \bibinfo{pages}{2272--2296}.
\newblock
\urldef\tempurl%
\url{https://doi.org/10.1002/spe.3340}
\showDOI{\tempurl}
\showeprint{https://onlinelibrary.wiley.com/doi/pdf/10.1002/spe.3340}


\bibitem[Barrios and Kumar(2023)]%
        {barrios23caching-edge}
\bibfield{author}{\bibinfo{person}{Carlos Barrios} {and} \bibinfo{person}{Mohan
  Kumar}.} \bibinfo{year}{2023}\natexlab{}.
\newblock \showarticletitle{Service Caching and Computation Reuse Strategies at
  the Edge: A Survey}.
\newblock \bibinfo{journal}{\emph{ACM Comput. Surv.}} \bibinfo{volume}{56},
  \bibinfo{number}{2}, Article \bibinfo{articleno}{43} (\bibinfo{date}{Sept.}
  \bibinfo{year}{2023}), \bibinfo{numpages}{38}~pages.
\newblock
\showISSN{0360-0300}
\urldef\tempurl%
\url{https://doi.org/10.1145/3609504}
\showDOI{\tempurl}


\bibitem[Batool et~al\mbox{.}(2026)]%
        {batool2026ai}
\bibfield{author}{\bibinfo{person}{Iqra Batool}, \bibinfo{person}{Mostafa~M.
  Fouda}, {and} \bibinfo{person}{Zubair~Md Fadlullah}.}
  \bibinfo{year}{2026}\natexlab{}.
\newblock \showarticletitle{AI-native predictive scheduling for computing power
  networks: A multi-scale learning framework for ultra-low latency 6G
  applications}.
\newblock \bibinfo{journal}{\emph{IEEE Transactions on Network Science and
  Engineering}}  \bibinfo{volume}{13} (\bibinfo{year}{2026}),
  \bibinfo{pages}{6193--6210}.
\newblock
\urldef\tempurl%
\url{https://doi.org/10.1109/TNSE.2026.3650877}
\showDOI{\tempurl}


\bibitem[Bhat et~al\mbox{.}(2026)]%
        {11434192}
\bibfield{author}{\bibinfo{person}{Shashidhar Bhat},
  \bibinfo{person}{Sathwik~Rao Sirikonda}, \bibinfo{person}{Vikas Katoch},
  {and} \bibinfo{person}{Rishabh Jain}.} \bibinfo{year}{2026}\natexlab{}.
\newblock \showarticletitle{Carbon-Kube: A Kubernetes-Native Framework for
  Multi-Objective Carbon-Aware Scheduling of Big Data Pipelines}. In
  \bibinfo{booktitle}{\emph{2026 9th International Conference on Electronics,
  Materials Engineering \& Nano-Technology (IEMENTech)}},
  Vol.~\bibinfo{volume}{9}. \bibinfo{pages}{1--6}.
\newblock
\urldef\tempurl%
\url{https://doi.org/10.1109/IEMENTech202669403.2026.11434192}
\showDOI{\tempurl}


\bibitem[Buyya et~al\mbox{.}(2024)]%
        {buyya2024energy}
\bibfield{author}{\bibinfo{person}{Rajkumar Buyya}, \bibinfo{person}{Shashikant
  Ilager}, {and} \bibinfo{person}{Patricia Arroba}.}
  \bibinfo{year}{2024}\natexlab{}.
\newblock \showarticletitle{Energy-efficiency and sustainability in new
  generation cloud computing: a vision and directions for integrated management
  of data centre resources and workloads}.
\newblock \bibinfo{journal}{\emph{Software: Practice and Experience}}
  \bibinfo{volume}{54}, \bibinfo{number}{1} (\bibinfo{year}{2024}),
  \bibinfo{pages}{24--38}.
\newblock
\urldef\tempurl%
\url{https://doi.org/10.1002/spe.3248}
\showDOI{\tempurl}
\showeprint{https://onlinelibrary.wiley.com/doi/pdf/10.1002/spe.3248}


\bibitem[Chien et~al\mbox{.}(2023)]%
        {chien23genai-rq-direct}
\bibfield{author}{\bibinfo{person}{Andrew~A Chien}, \bibinfo{person}{Liuzixuan
  Lin}, \bibinfo{person}{Hai Nguyen}, \bibinfo{person}{Varsha Rao},
  \bibinfo{person}{Tristan Sharma}, {and} \bibinfo{person}{Rajini
  Wijayawardana}.} \bibinfo{year}{2023}\natexlab{}.
\newblock \showarticletitle{Reducing the Carbon Impact of Generative AI
  Inference (today and in 2035)}. In \bibinfo{booktitle}{\emph{Proceedings of
  the 2nd Workshop on Sustainable Computer Systems}}. Article
  \bibinfo{articleno}{11}, \bibinfo{numpages}{7}~pages.
\newblock


\bibitem[Farahani et~al\mbox{.}(2026)]%
        {farahani2024towards}
\bibfield{author}{\bibinfo{person}{Reza Farahani}, \bibinfo{person}{Zoha
  Azimi~Ourimi}, \bibinfo{person}{Christian Timmerer}, {and}
  \bibinfo{person}{Radu Prodan}.} \bibinfo{year}{2026}\natexlab{}.
\newblock \showarticletitle{Towards AI-Assisted Sustainable Adaptive Video
  Streaming Systems: Tutorial and Survey}.
\newblock \bibinfo{journal}{\emph{ACM Comput. Surv.}} \bibinfo{volume}{58},
  \bibinfo{number}{15}, Article \bibinfo{articleno}{378} (\bibinfo{date}{Aug.}
  \bibinfo{year}{2026}), \bibinfo{numpages}{37}~pages.
\newblock
\showISSN{0360-0300}
\urldef\tempurl%
\url{https://doi.org/10.1145/3838182}
\showDOI{\tempurl}


\bibitem[Gao et~al\mbox{.}(2026)]%
        {11407972}
\bibfield{author}{\bibinfo{person}{Juncai Gao}, \bibinfo{person}{Zhuoyue Chen},
  \bibinfo{person}{Zhihao Hu}, \bibinfo{person}{Zhanqi Cui},
  \bibinfo{person}{Ying Chen}, {and} \bibinfo{person}{Jiwei Huang}.}
  \bibinfo{year}{2026}\natexlab{}.
\newblock \showarticletitle{Carbon-Aware Dynamic Task Scheduling in
  Hierarchical Cloud–Edge Systems for IoT Devices}.
\newblock \bibinfo{journal}{\emph{IEEE Internet of Things Journal}}
  \bibinfo{volume}{13}, \bibinfo{number}{10} (\bibinfo{year}{2026}),
  \bibinfo{pages}{20883--20896}.
\newblock
\urldef\tempurl%
\url{https://doi.org/10.1109/JIOT.2026.3665092}
\showDOI{\tempurl}


\bibitem[Google(2022)]%
        {google22spot-vms}
\bibfield{author}{\bibinfo{person}{Google}.} \bibinfo{year}{2022}\natexlab{}.
\newblock \bibinfo{title}{Rethinking your VM strategy with Spot VMs}.
\newblock
\newblock
\urldef\tempurl%
\url{https://cloud.google.com/blog/topics/cost-management/rethinking-your-vm-strategy-spot-vms}
\showURL{%
\tempurl}


\bibitem[Goudarzi et~al\mbox{.}(2022)]%
        {goudarzi22edge-iot}
\bibfield{author}{\bibinfo{person}{Mohammad Goudarzi},
  \bibinfo{person}{Marimuthu Palaniswami}, {and} \bibinfo{person}{Rajkumar
  Buyya}.} \bibinfo{year}{2022}\natexlab{}.
\newblock \showarticletitle{Scheduling IoT Applications in Edge and Fog
  Computing Environments: A Taxonomy and Future Directions}.
\newblock \bibinfo{journal}{\emph{ACM Comput. Surv.}} \bibinfo{volume}{55},
  \bibinfo{number}{7}, Article \bibinfo{articleno}{152} (\bibinfo{date}{Dec.}
  \bibinfo{year}{2022}), \bibinfo{numpages}{41}~pages.
\newblock
\showISSN{0360-0300}
\urldef\tempurl%
\url{https://doi.org/10.1145/3544836}
\showDOI{\tempurl}


\bibitem[Govindan et~al\mbox{.}(2011)]%
        {govindan11stored-energy-dc}
\bibfield{author}{\bibinfo{person}{Sriram Govindan}, \bibinfo{person}{Anand
  Sivasubramaniam}, {and} \bibinfo{person}{Bhuvan Urgaonkar}.}
  \bibinfo{year}{2011}\natexlab{}.
\newblock \showarticletitle{Benefits and limitations of tapping into stored
  energy for datacenters}.
\newblock   \bibinfo{volume}{39} (\bibinfo{date}{June} \bibinfo{year}{2011}),
  \bibinfo{pages}{341–352}.
\newblock


\bibitem[Gsteiger et~al\mbox{.}(2024)]%
        {gsteiger24caribou}
\bibfield{author}{\bibinfo{person}{Viktor~Urban Gsteiger}, \bibinfo{person}{Pin
  Hong~(Daniel) Long}, \bibinfo{person}{Yiran~(Jerry) Sun},
  \bibinfo{person}{Parshan Javanrood}, {and} \bibinfo{person}{Mohammad
  Shahrad}.} \bibinfo{year}{2024}\natexlab{}.
\newblock \showarticletitle{Caribou: Fine-Grained Geospatial Shifting of
  Serverless Applications for Sustainability}. In
  \bibinfo{booktitle}{\emph{Proceedings of the ACM SIGOPS 30th Symposium on
  Operating Systems Principles}}. \bibinfo{pages}{403–420}.
\newblock


\bibitem[Gupta et~al\mbox{.}(2024)]%
        {gupta24meta-embodied}
\bibfield{author}{\bibinfo{person}{Nishant Gupta}, \bibinfo{person}{Iyswarya
  Narayanan}, \bibinfo{person}{Shivam Handa}, \bibinfo{person}{Sayak
  Chakraborti}, \bibinfo{person}{Pankit Thapar}, \bibinfo{person}{Baohua Shan},
  \bibinfo{person}{Ariel Rao}, \bibinfo{person}{Yuanlai Liu},
  \bibinfo{person}{Pengyuan Wang}, \bibinfo{person}{Yuqing Wu},
  \bibinfo{person}{Qingyi Gao}, \bibinfo{person}{Chris Chao-Chun Cheng},
  \bibinfo{person}{Sihan You}, \bibinfo{person}{Louis Huang},
  \bibinfo{person}{Jingyuan Fan}, \bibinfo{person}{Kenny Yu},
  \bibinfo{person}{Kevin Lin}, \bibinfo{person}{Tengfei Mu},
  \bibinfo{person}{Parth Malani}, \bibinfo{person}{Haiying Wang},
  \bibinfo{person}{Trey Lu}, {and} \bibinfo{person}{Peter Zhang}.}
  \bibinfo{year}{2024}\natexlab{}.
\newblock \showarticletitle{Dynamic Idle Resource Leasing To Safely
  Oversubscribe Capacity At Meta}. In \bibinfo{booktitle}{\emph{Proceedings of
  the 2024 ACM Symposium on Cloud Computing}}. \bibinfo{pages}{792–810}.
\newblock


\bibitem[Gupta et~al\mbox{.}(2022)]%
        {gupta22act}
\bibfield{author}{\bibinfo{person}{Udit Gupta}, \bibinfo{person}{Mariam
  Elgamal}, \bibinfo{person}{Gage Hills}, \bibinfo{person}{Gu-Yeon Wei},
  \bibinfo{person}{Hsien-Hsin~S. Lee}, \bibinfo{person}{David Brooks}, {and}
  \bibinfo{person}{Carole-Jean Wu}.} \bibinfo{year}{2022}\natexlab{}.
\newblock \showarticletitle{ACT: designing sustainable computer systems with an
  architectural carbon modeling tool}. In \bibinfo{booktitle}{\emph{Proceedings
  of the 49th Annual International Symposium on Computer Architecture}}.
  \bibinfo{pages}{784–799}.
\newblock


\bibitem[Jahanshahi et~al\mbox{.}(2022)]%
        {jahanshahi2022powermorph}
\bibfield{author}{\bibinfo{person}{Ali Jahanshahi}, \bibinfo{person}{Nanpeng
  Yu}, {and} \bibinfo{person}{Daniel Wong}.} \bibinfo{year}{2022}\natexlab{}.
\newblock \showarticletitle{PowerMorph: QoS-Aware Server Power Reshaping for
  Data Center Regulation Service}.
\newblock \bibinfo{journal}{\emph{ACM Trans. Archit. Code Optim.}}
  \bibinfo{volume}{19}, Article \bibinfo{articleno}{36} (\bibinfo{year}{2022}).
\newblock


\bibitem[Jeyaraj et~al\mbox{.}(2023)]%
        {jeyaraj23em-cloud-iot}
\bibfield{author}{\bibinfo{person}{Rathinaraja Jeyaraj},
  \bibinfo{person}{Anandkumar Balasubramaniam}, \bibinfo{person}{Ajay~Kumara
  M.A.}, \bibinfo{person}{Nadra Guizani}, {and} \bibinfo{person}{Anand Paul}.}
  \bibinfo{year}{2023}\natexlab{}.
\newblock \showarticletitle{Resource Management in Cloud and Cloud-influenced
  Technologies for Internet of Things Applications}.
\newblock \bibinfo{journal}{\emph{ACM Comput. Surv.}} \bibinfo{volume}{55},
  \bibinfo{number}{12}, Article \bibinfo{articleno}{242} (\bibinfo{date}{March}
  \bibinfo{year}{2023}), \bibinfo{numpages}{37}~pages.
\newblock
\showISSN{0360-0300}
\urldef\tempurl%
\url{https://doi.org/10.1145/3571729}
\showDOI{\tempurl}


\bibitem[Ji et~al\mbox{.}(2024)]%
        {ji24scarif-embodied-design}
\bibfield{author}{\bibinfo{person}{Shixin Ji}, \bibinfo{person}{Zhuoping Yang},
  \bibinfo{person}{Xingzhen Chen}, \bibinfo{person}{Stephen Cahoon},
  \bibinfo{person}{Jingtong Hu}, \bibinfo{person}{Yiyu Shi},
  \bibinfo{person}{Alex~K. Jones}, {and} \bibinfo{person}{Peipei Zhou}.}
  \bibinfo{year}{2024}\natexlab{}.
\newblock \showarticletitle{SCARIF: Towards Carbon Modeling of Cloud Servers
  with Accelerators}. In \bibinfo{booktitle}{\emph{Proceedings of the 2024 IEEE
  Computer Society Annual Symposium on VLSI (ISVLSI)}}.
  \bibinfo{pages}{496--501}.
\newblock


\bibitem[Jiang et~al\mbox{.}(2026)]%
        {11666925}
\bibfield{author}{\bibinfo{person}{Chunmao Jiang}, \bibinfo{person}{Gege Li},
  \bibinfo{person}{Ruyi Ye}, {and} \bibinfo{person}{Hao Zhang}.}
  \bibinfo{year}{2026}\natexlab{}.
\newblock \showarticletitle{CARBONMOBILE: Carbon-Efficient Multi-Edge Workload
  Orchestration for Mobile IoT}.
\newblock \bibinfo{journal}{\emph{IEEE Internet of Things Journal}}
  (\bibinfo{year}{2026}), \bibinfo{pages}{1--1}.
\newblock
\urldef\tempurl%
\url{https://doi.org/10.1109/JIOT.2026.3727208}
\showDOI{\tempurl}


\bibitem[Kaur et~al\mbox{.}(2020)]%
        {kaur20keids-rt}
\bibfield{author}{\bibinfo{person}{Kuljeet Kaur}, \bibinfo{person}{Sahil Garg},
  \bibinfo{person}{Georges Kaddoum}, \bibinfo{person}{Syed~Hassan Ahmed}, {and}
  \bibinfo{person}{Mohammed Atiquzzaman}.} \bibinfo{year}{2020}\natexlab{}.
\newblock \showarticletitle{KEIDS: Kubernetes-Based Energy and Interference
  Driven Scheduler for Industrial IoT in Edge-Cloud Ecosystem}.
\newblock \bibinfo{journal}{\emph{IEEE Internet of Things Journal}}
  \bibinfo{volume}{7}, \bibinfo{number}{5} (\bibinfo{year}{2020}),
  \bibinfo{pages}{4228--4237}.
\newblock
\urldef\tempurl%
\url{https://doi.org/10.1109/JIOT.2019.2939534}
\showDOI{\tempurl}


\bibitem[Kong and Liu(2014)]%
        {kong14green-energy-dc}
\bibfield{author}{\bibinfo{person}{Fanxin Kong} {and} \bibinfo{person}{Xue
  Liu}.} \bibinfo{year}{2014}\natexlab{}.
\newblock \showarticletitle{A Survey on Green-Energy-Aware Power Management for
  Datacenters}.
\newblock  \bibinfo{volume}{47}, \bibinfo{number}{2}, Article
  \bibinfo{articleno}{30} (\bibinfo{date}{Nov.} \bibinfo{year}{2014}),
  \bibinfo{numpages}{38}~pages.
\newblock
\showISSN{0360-0300}
\urldef\tempurl%
\url{https://doi.org/10.1145/2642708}
\showURL{%
\tempurl}


\bibitem[Li et~al\mbox{.}(2013)]%
        {li2013chameleon}
\bibfield{author}{\bibinfo{person}{Chao Li}, \bibinfo{person}{Xian Li},
  \bibinfo{person}{Rui Wang}, \bibinfo{person}{Tao Li},
  \bibinfo{person}{Nilanjan Goswami}, {and} \bibinfo{person}{Depei Qian}.}
  \bibinfo{year}{2013}\natexlab{}.
\newblock \showarticletitle{Chameleon: Adapting throughput server to
  time-varying green power budget using online learning}. In
  \bibinfo{booktitle}{\emph{Proceedings of the International Symposium on Low
  Power Electronics and Design (ISLPED)}}. \bibinfo{pages}{100--105}.
\newblock


\bibitem[Liu et~al\mbox{.}(2026)]%
        {liu2026edge}
\bibfield{author}{\bibinfo{person}{Jing Liu}, \bibinfo{person}{Yao Du},
  \bibinfo{person}{Kun Yang}, \bibinfo{person}{Jiaqi Wu}, \bibinfo{person}{Yan
  Wang}, \bibinfo{person}{Xiping Hu}, \bibinfo{person}{Zehua Wang},
  \bibinfo{person}{Yang Liu}, \bibinfo{person}{Peng Sun},
  \bibinfo{person}{Azzedine Boukerche}, {and} \bibinfo{person}{Victor C.~M.
  Leung}.} \bibinfo{year}{2026}\natexlab{}.
\newblock \showarticletitle{Edge-cloud collaborative computing on distributed
  intelligence and model optimization: A survey}.
\newblock \bibinfo{journal}{\emph{IEEE Communications Surveys \& Tutorials}}
  \bibinfo{volume}{28} (\bibinfo{year}{2026}), \bibinfo{pages}{5049--5080}.
\newblock
\urldef\tempurl%
\url{https://doi.org/10.1109/COMST.2026.3669216}
\showDOI{\tempurl}


\bibitem[Liu and Buyya(2020)]%
        {liu20strem-processing}
\bibfield{author}{\bibinfo{person}{Xunyun Liu} {and} \bibinfo{person}{Rajkumar
  Buyya}.} \bibinfo{year}{2020}\natexlab{}.
\newblock \showarticletitle{Resource Management and Scheduling in Distributed
  Stream Processing Systems: A Taxonomy, Review, and Future Directions}.
\newblock \bibinfo{journal}{\emph{ACM Comput. Surv.}} \bibinfo{volume}{53},
  \bibinfo{number}{3}, Article \bibinfo{articleno}{50} (\bibinfo{date}{May}
  \bibinfo{year}{2020}), \bibinfo{numpages}{41}~pages.
\newblock
\showISSN{0360-0300}
\urldef\tempurl%
\url{https://doi.org/10.1145/3355399}
\showDOI{\tempurl}


\bibitem[McAllister et~al\mbox{.}(2024)]%
        {sara24wrencache}
\bibfield{author}{\bibinfo{person}{Sara McAllister},
  \bibinfo{person}{Yucong~"Sherry" Wang}, \bibinfo{person}{Benjamin Berg},
  \bibinfo{person}{Daniel~S. Berger}, \bibinfo{person}{George Amvrosiadis},
  \bibinfo{person}{Nathan Beckmann}, {and} \bibinfo{person}{Gregory~R.
  Ganger}.} \bibinfo{year}{2024}\natexlab{}.
\newblock \showarticletitle{{FairyWREN}: A Sustainable Cache for Emerging
  {Write-Read-Erase} Flash Interfaces}. In
  \bibinfo{booktitle}{\emph{Proceedings of the 18th USENIX Symposium on
  Operating Systems Design and Implementation (OSDI 24)}}.
  \bibinfo{pages}{745--764}.
\newblock


\bibitem[Murillo et~al\mbox{.}(2024)]%
        {murillo24cdnshifter}
\bibfield{author}{\bibinfo{person}{Jorge Murillo}, \bibinfo{person}{Walid~A.
  Hanafy}, \bibinfo{person}{David Irwin}, \bibinfo{person}{Ramesh Sitaraman},
  {and} \bibinfo{person}{Prashant Shenoy}.} \bibinfo{year}{2024}\natexlab{}.
\newblock \showarticletitle{CDN-Shifter: Leveraging Spatial Workload Shifting
  to Decarbonize Content Delivery Networks}. In
  \bibinfo{booktitle}{\emph{Proceedings of the 2024 ACM Symposium on Cloud
  Computing}}. \bibinfo{pages}{505–521}.
\newblock


\bibitem[Queiroz et~al\mbox{.}(2023)]%
        {queiroz23cont-virt-rt}
\bibfield{author}{\bibinfo{person}{Rui Queiroz}, \bibinfo{person}{Tiago Cruz},
  \bibinfo{person}{J\'{e}r\^{o}me Mendes}, \bibinfo{person}{Pedro Sousa}, {and}
  \bibinfo{person}{Paulo Sim\~{o}es}.} \bibinfo{year}{2023}\natexlab{}.
\newblock \showarticletitle{Container-based Virtualization for Real-time
  Industrial Systems—A Systematic Review}.
\newblock \bibinfo{journal}{\emph{ACM Comput. Surv.}} \bibinfo{volume}{56},
  \bibinfo{number}{3}, Article \bibinfo{articleno}{59} (\bibinfo{date}{Oct.}
  \bibinfo{year}{2023}), \bibinfo{numpages}{38}~pages.
\newblock
\showISSN{0360-0300}
\urldef\tempurl%
\url{https://doi.org/10.1145/3617591}
\showDOI{\tempurl}


\bibitem[Radu(2017)]%
        {radu17green-cloud-computing}
\bibfield{author}{\bibinfo{person}{Laura-Diana Radu}.}
  \bibinfo{year}{2017}\natexlab{}.
\newblock \showarticletitle{Green Cloud Computing: A Literature Survey}.
\newblock \bibinfo{journal}{\emph{Symmetry}} \bibinfo{volume}{9},
  \bibinfo{number}{12} (\bibinfo{year}{2017}), \bibinfo{numpages}{20}~pages.
\newblock
\showISSN{2073-8994}
\urldef\tempurl%
\url{https://doi.org/10.3390/sym9120295}
\showDOI{\tempurl}


\bibitem[Rasouli et~al\mbox{.}(2025)]%
        {rasouli2025resource}
\bibfield{author}{\bibinfo{person}{Nayereh Rasouli}, \bibinfo{person}{Cristian
  Klein}, {and} \bibinfo{person}{Erik Elmroth}.}
  \bibinfo{year}{2025}\natexlab{}.
\newblock \showarticletitle{Resource Management for Mission-Critical
  Applications in Edge Computing: Systematic Review on Recent Research and Open
  Issues}.
\newblock \bibinfo{journal}{\emph{ACM Comput. Surv.}} \bibinfo{volume}{58},
  \bibinfo{number}{3}, Article \bibinfo{articleno}{71} (\bibinfo{date}{Sept.}
  \bibinfo{year}{2025}), \bibinfo{numpages}{37}~pages.
\newblock
\showISSN{0360-0300}
\urldef\tempurl%
\url{https://doi.org/10.1145/3762181}
\showDOI{\tempurl}


\bibitem[Sajid et~al\mbox{.}(2021)]%
        {sajid2021blockchain}
\bibfield{author}{\bibinfo{person}{Sara Sajid}, \bibinfo{person}{Muhammad
  Jawad}, \bibinfo{person}{Kanza Hamid}, \bibinfo{person}{Muhammad~U.S. Khan},
  \bibinfo{person}{Sahibzada~M. Ali}, \bibinfo{person}{Assad Abbas}, {and}
  \bibinfo{person}{Samee~U. Khan}.} \bibinfo{year}{2021}\natexlab{}.
\newblock \showarticletitle{Blockchain-based decentralized workload and energy
  management of geo-distributed data centers}.
\newblock \bibinfo{journal}{\emph{Sustainable Computing: Informatics and
  Systems}}  \bibinfo{volume}{29} (\bibinfo{year}{2021}),
  \bibinfo{pages}{100--461}.
\newblock


\bibitem[Salaht et~al\mbox{.}(2020)]%
        {salaht20placement-edge}
\bibfield{author}{\bibinfo{person}{Farah~A\"{\i}t Salaht},
  \bibinfo{person}{Fr\'{e}d\'{e}ric Desprez}, {and} \bibinfo{person}{Adrien
  Lebre}.} \bibinfo{year}{2020}\natexlab{}.
\newblock \showarticletitle{An Overview of Service Placement Problem in Fog and
  Edge Computing}.
\newblock \bibinfo{journal}{\emph{ACM Comput. Surv.}} \bibinfo{volume}{53},
  \bibinfo{number}{3}, Article \bibinfo{articleno}{65} (\bibinfo{date}{June}
  \bibinfo{year}{2020}), \bibinfo{numpages}{35}~pages.
\newblock
\showISSN{0360-0300}
\urldef\tempurl%
\url{https://doi.org/10.1145/3391196}
\showDOI{\tempurl}


\bibitem[Shandilya(2020)]%
        {microsoft20spot-vms}
\bibfield{author}{\bibinfo{person}{Varun Shandilya}.}
  \bibinfo{year}{2020}\natexlab{}.
\newblock \bibinfo{title}{Announcing the general availability of Azure Spot
  Virtual Machines}.
\newblock
\newblock
\urldef\tempurl%
\url{https://azure.microsoft.com/en-us/blog/announcing-the-general-availability-of-azure-spot-virtual-machines/}
\showURL{%
\tempurl}


\bibitem[Sharma et~al\mbox{.}(2011)]%
        {sharma11blink}
\bibfield{author}{\bibinfo{person}{Navin Sharma}, \bibinfo{person}{Sean
  Barker}, \bibinfo{person}{David Irwin}, {and} \bibinfo{person}{Prashant
  Shenoy}.} \bibinfo{year}{2011}\natexlab{}.
\newblock \showarticletitle{Blink: managing server clusters on intermittent
  power}. In \bibinfo{booktitle}{\emph{Proceedings of the Sixteenth
  International Conference on Architectural Support for Programming Languages
  and Operating Systems}}. \bibinfo{pages}{185–198}.
\newblock


\bibitem[Shuja et~al\mbox{.}(2016)]%
        {junaid16sus-cdc}
\bibfield{author}{\bibinfo{person}{Junaid Shuja}, \bibinfo{person}{Abdullah
  Gani}, \bibinfo{person}{Shahaboddin Shamshirband},
  \bibinfo{person}{Raja~Wasim Ahmad}, {and} \bibinfo{person}{Kashif Bilal}.}
  \bibinfo{year}{2016}\natexlab{}.
\newblock \showarticletitle{Sustainable Cloud Data Centers: A survey of
  enabling techniques and technologies}.
\newblock \bibinfo{journal}{\emph{Renewable and Sustainable Energy Reviews}}
  \bibinfo{volume}{62} (\bibinfo{year}{2016}), \bibinfo{pages}{195--214}.
\newblock
\showISSN{1364-0321}
\urldef\tempurl%
\url{https://doi.org/10.1016/j.rser.2016.04.034}
\showDOI{\tempurl}


\bibitem[Souza et~al\mbox{.}(2023)]%
        {souza23ecovisor}
\bibfield{author}{\bibinfo{person}{Abel Souza}, \bibinfo{person}{Noman Bashir},
  \bibinfo{person}{Jorge Murillo}, \bibinfo{person}{Walid Hanafy},
  \bibinfo{person}{Qianlin Liang}, \bibinfo{person}{David Irwin}, {and}
  \bibinfo{person}{Prashant Shenoy}.} \bibinfo{year}{2023}\natexlab{}.
\newblock \showarticletitle{Ecovisor: A Virtual Energy System for
  Carbon-Efficient Applications}. In \bibinfo{booktitle}{\emph{Proceedings of
  the 28th ACM International Conference on Architectural Support for
  Programming Languages and Operating Systems, Volume 2}}.
  \bibinfo{pages}{252–265}.
\newblock


\bibitem[Souza et~al\mbox{.}(2024b)]%
        {souza24casper}
\bibfield{author}{\bibinfo{person}{Abel Souza}, \bibinfo{person}{Shruti
  Jasoria}, \bibinfo{person}{Basundhara Chakrabarty},
  \bibinfo{person}{Alexander Bridgwater}, \bibinfo{person}{Axel Lundberg},
  \bibinfo{person}{Filip Skogh}, \bibinfo{person}{Ahmed Ali-Eldin},
  \bibinfo{person}{David Irwin}, {and} \bibinfo{person}{Prashant Shenoy}.}
  \bibinfo{year}{2024}\natexlab{b}.
\newblock \showarticletitle{CASPER: Carbon-Aware Scheduling and Provisioning
  for Distributed Web Services}. In \bibinfo{booktitle}{\emph{Proceedings of
  the 14th International Green and Sustainable Computing Conference}}.
  \bibinfo{pages}{67–73}.
\newblock


\bibitem[Souza et~al\mbox{.}(2024a)]%
        {souza24maintanance-iot}
\bibfield{author}{\bibinfo{person}{Paulo Souza}, \bibinfo{person}{Tiago
  Ferreto}, {and} \bibinfo{person}{Rodrigo Calheiros}.}
  \bibinfo{year}{2024}\natexlab{a}.
\newblock \showarticletitle{Maintenance Operations on Cloud, Edge, and IoT
  Environments: Taxonomy, Survey, and Research Challenges}.
\newblock \bibinfo{journal}{\emph{ACM Comput. Surv.}} \bibinfo{volume}{56},
  \bibinfo{number}{10}, Article \bibinfo{articleno}{256} (\bibinfo{date}{June}
  \bibinfo{year}{2024}), \bibinfo{numpages}{38}~pages.
\newblock
\showISSN{0360-0300}
\urldef\tempurl%
\url{https://doi.org/10.1145/3659097}
\showDOI{\tempurl}


\bibitem[Sukprasert et~al\mbox{.}(2024)]%
        {sukprasert24limits-of-load-shift}
\bibfield{author}{\bibinfo{person}{Thanathorn Sukprasert},
  \bibinfo{person}{Abel Souza}, \bibinfo{person}{Noman Bashir},
  \bibinfo{person}{David Irwin}, {and} \bibinfo{person}{Prashant Shenoy}.}
  \bibinfo{year}{2024}\natexlab{}.
\newblock \showarticletitle{On the Limitations of Carbon-Aware Temporal and
  Spatial Workload Shifting in the Cloud}. In
  \bibinfo{booktitle}{\emph{Proceedings of the Nineteenth European Conference
  on Computer Systems}}. \bibinfo{pages}{924–941}.
\newblock


\bibitem[Sun et~al\mbox{.}(2026)]%
        {11619171}
\bibfield{author}{\bibinfo{person}{Bowen Sun}, \bibinfo{person}{Christos
  Antonopoulos}, \bibinfo{person}{Evgenia Smirni}, \bibinfo{person}{Bin Ren},
  \bibinfo{person}{Nikolaos Bellas}, {and} \bibinfo{person}{Spyros Lalis}.}
  \bibinfo{year}{2026}\natexlab{}.
\newblock \showarticletitle{Green or Fast? Learning to Balance Cold Starts and
  Idle Carbon in Serverless Computing}. In \bibinfo{booktitle}{\emph{2026 IEEE
  26th International Symposium on Cluster, Cloud and Internet Computing
  (CCGrid)}}. \bibinfo{pages}{254--264}.
\newblock
\urldef\tempurl%
\url{https://doi.org/10.1109/CCGrid68966.2026.00034}
\showDOI{\tempurl}


\bibitem[Sun et~al\mbox{.}(2024)]%
        {sun24modular-datacenters}
\bibfield{author}{\bibinfo{person}{Jinghan Sun}, \bibinfo{person}{Zibo Gong},
  \bibinfo{person}{Anup Agarwal}, \bibinfo{person}{Shadi Noghabi},
  \bibinfo{person}{Ranveer Chandra}, \bibinfo{person}{Marc Snir}, {and}
  \bibinfo{person}{Jian Huang}.} \bibinfo{year}{2024}\natexlab{}.
\newblock \showarticletitle{Exploring the Efficiency of Renewable Energy-based
  Modular Data Centers at Scale}. In \bibinfo{booktitle}{\emph{Proceedings of
  the 2024 ACM Symposium on Cloud Computing}}. \bibinfo{pages}{552–569}.
\newblock


\bibitem[Tannu and Nair(2023)]%
        {tannu23dirty-ssd}
\bibfield{author}{\bibinfo{person}{Swamit Tannu} {and}
  \bibinfo{person}{Prashant~J. Nair}.} \bibinfo{year}{2023}\natexlab{}.
\newblock \showarticletitle{The Dirty Secret of SSDs: Embodied Carbon}.
\newblock \bibinfo{journal}{\emph{SIGENERGY Energy Inform. Rev.}}
  \bibinfo{volume}{3} (\bibinfo{year}{2023}), \bibinfo{pages}{4–9}.
\newblock


\bibitem[Wang et~al\mbox{.}(2024)]%
        {wang24greensku}
\bibfield{author}{\bibinfo{person}{Jaylen Wang}, \bibinfo{person}{Daniel~S.
  Berger}, \bibinfo{person}{Fiodar Kazhamiaka}, \bibinfo{person}{Celine
  Irvene}, \bibinfo{person}{Chaojie Zhang}, \bibinfo{person}{Esha Choukse},
  \bibinfo{person}{Kali Frost}, \bibinfo{person}{Rodrigo Fonseca},
  \bibinfo{person}{Brijesh Warrier}, \bibinfo{person}{Chetan Bansal},
  \bibinfo{person}{Jonathan Stern}, \bibinfo{person}{Ricardo Bianchini}, {and}
  \bibinfo{person}{Akshitha Sriraman}.} \bibinfo{year}{2024}\natexlab{}.
\newblock \showarticletitle{Designing Cloud Servers for Lower Carbon}. In
  \bibinfo{booktitle}{\emph{Proceedings of the 2024 ACM/IEEE 51st Annual
  International Symposium on Computer Architecture (ISCA)}}.
  \bibinfo{pages}{452--470}.
\newblock


\bibitem[Wang et~al\mbox{.}(2023)]%
        {wang23peeling-carbon}
\bibfield{author}{\bibinfo{person}{Jaylen Wang}, \bibinfo{person}{Udit Gupta},
  {and} \bibinfo{person}{Akshitha Sriraman}.} \bibinfo{year}{2023}\natexlab{}.
\newblock \showarticletitle{Peeling Back the Carbon Curtain: Carbon
  Optimization Challenges in Cloud Computing}. In
  \bibinfo{booktitle}{\emph{Proceedings of the 2nd Workshop on Sustainable
  Computer Systems}}. Article \bibinfo{articleno}{8},
  \bibinfo{numpages}{7}~pages.
\newblock


\bibitem[Yuan et~al\mbox{.}(2019)]%
        {yuan2019delaytolerantgeodatacenters}
\bibfield{author}{\bibinfo{person}{Haitao Yuan}, \bibinfo{person}{Jing Bi},
  {and} \bibinfo{person}{MengChu Zhou}.} \bibinfo{year}{2019}\natexlab{}.
\newblock \showarticletitle{Spatiotemporal Task Scheduling for Heterogeneous
  Delay-Tolerant Applications in Distributed Green Data Centers}.
\newblock \bibinfo{journal}{\emph{IEEE Transactions on Automation Science and
  Engineering}}  \bibinfo{volume}{16} (\bibinfo{year}{2019}),
  \bibinfo{pages}{1686--1697}.
\newblock


\bibitem[Zhao et~al\mbox{.}(2023)]%
        {zhao2023unsustainableaffinity}
\bibfield{author}{\bibinfo{person}{Jiechen Zhao}, \bibinfo{person}{Katie Lim},
  \bibinfo{person}{Thomas Anderson}, {and} \bibinfo{person}{Natalie
  Enright~Jerger}.} \bibinfo{year}{2023}\natexlab{}.
\newblock \showarticletitle{The Case of Unsustainable CPU Affinity}. In
  \bibinfo{booktitle}{\emph{Proceedings of the 2nd Workshop on Sustainable
  Computer Systems}}. Article \bibinfo{articleno}{1},
  \bibinfo{numpages}{7}~pages.
\newblock


\bibitem[Zhong et~al\mbox{.}(2024)]%
        {yuhong24cxl-memory}
\bibfield{author}{\bibinfo{person}{Yuhong Zhong}, \bibinfo{person}{Daniel~S.
  Berger}, \bibinfo{person}{Carl Waldspurger}, \bibinfo{person}{Ryan Wee},
  \bibinfo{person}{Ishwar Agarwal}, \bibinfo{person}{Rajat Agarwal},
  \bibinfo{person}{Frank Hady}, \bibinfo{person}{Karthik Kumar},
  \bibinfo{person}{Mark~D. Hill}, \bibinfo{person}{Mosharaf Chowdhury}, {and}
  \bibinfo{person}{Asaf Cidon}.} \bibinfo{year}{2024}\natexlab{}.
\newblock \showarticletitle{Managing Memory Tiers with {CXL} in Virtualized
  Environments}. In \bibinfo{booktitle}{\emph{Proceedings of the 18th USENIX
  Symposium on Operating Systems Design and Implementation (OSDI 24)}}.
  \bibinfo{pages}{37--56}.
\newblock


\end{thebibliography}

\end{document}